\documentclass[]{emulateapj}\usepackage{psfig,graphicx,natbib, gensymb}

\begin{document}
\title{Modelling the Rossiter-McLaughlin Effect: Impact of the Convective Centre-to-Limb Variations in the Stellar Photosphere}
\author{H.~M. Cegla\altaffilmark{1}, M. Oshagh\altaffilmark{2}, C.~A. Watson\altaffilmark{1}, P. Figueira\altaffilmark{2}, N.~C. Santos\altaffilmark{2,3}, S. Shelyag\altaffilmark{4}}
\altaffiltext{1}{Astrophysics Research Centre, School of Mathematics \& Physics, Queen's University Belfast, University Road, Belfast BT7 1NN, UK; {\tt h.cegla@qub.ac.uk}}
\altaffiltext{2}{Instituto de Astrof\' isica e Ci\^encias do Espa\c{c}o, Universidade do Porto, CAUP, Rua das Estrelas, PT4150-762 Porto, Portugal, PT}
\altaffiltext{3}{Departamento de F\'isica e Astronomia, Faculdade de Ci\^encias, Universidade do Porto, Rua Campo Alegre, 4169-007 Porto, Portugal}
\altaffiltext{4}{Monash Centre for Astrophysics, School of  Mathematical Sciences, Monash University, Clayton, Victoria,  3800, AU}

\begin{abstract}
Observations of the Rossiter-McLaughlin (RM) effect provide information on star-planet alignments, which can inform planetary migration and evolution theories. Here, we go beyond the classical RM modelling and explore the impact of a convective blueshift that varies across the stellar disc and non-Gaussian stellar photospheric profiles. We simulated an aligned hot Jupiter with a 4 d orbit about a Sun-like star and injected centre-to-limb velocity (and profile shape) variations based on radiative 3D magnetohydrodynamic simulations of solar surface convection. The residuals between our modelling and classical RM modelling were dependent on the intrinsic profile width and $v \sin i$; the amplitude of the residuals increased with increasing $v \sin i$, and with decreasing intrinsic profile width. For slowly rotating stars the centre-to-limb convective variation dominated the residuals (with amplitudes of 10s of cm~s$^{-1}$ to $\sim$1~m~s$^{-1}$); however, for faster rotating stars the dominant residual signature was due a non-Gaussian intrinsic profile (with amplitudes from 0.5-9~m~s$^{-1}$). When the impact factor was 0, neglecting to account for the convective centre-to-limb variation led to an uncertainty in the obliquity of $\sim$10-20$\degree$, even though the true $v \sin i$ was known. Additionally, neglecting to properly model an asymmetric intrinsic profile had a greater impact for more rapidly rotating stars (e.g. $v \sin i$ = 6~km~s$^{-1}$), and caused systematic errors on the order of $\sim$20$\degree$ in the measured obliquities. Hence, neglecting the impact of stellar surface convection may bias star-planet alignment measurements and consequently also theories on planetary migration and evolution.

\end{abstract}


\section{Introduction}
\label{sec:intro}
Radial velocity (RV) precision is primarily limited by instrumentation and our understanding of stellar spectral lines. Consequently, the continued improvement in instrumental precision demands an evermore accurate treatment of spectral line behaviour. This is clearly evident now as current spectrographs, such as HARPS, can routinely offer a precision of $\sim$~0.5 m~s$^{-1}$, while astrophysical phenomena can distort stellar lines and induce spurious velocity shifts ranging from several 10s of cm~s$^{-1}$ to 100s of m~s$^{-1}$ for solar-type stars (due to, for example, variations in gravitational redshift, stellar surface (magneto-)convection, natural oscillations, meridional circulation, spots, plages, and the attenuation of convective blueshift surrounding regions of high magnetic field; \citealt{cegla12,saar97,schrijver00,dumusque11a,beckers07,dumusque11b, boisse11,meunier13}.) 

Additionally, it is clear that the need for an accurate description of even low-amplitude phenomena will only intensify as spectrographs, such as ESPRESSO \citep{pepe14}, promise precisions of 10 cm~s$^{-1}$ or better by as early as 2017. Such astrophysical phenomena affect any high precision RV study. Spectroscopic observations of exoplanets are particularly affected by these phenomena as it can be extremely difficult to disentangle planetary and stellar signals from one another. This is in addition to the fact that stellar signals can masquerade as planetary signals \citep[e.g.][]{queloz01,desidera04,huelamo08,figueria10,santos14,robertson15}. 

Furthermore, ignoring certain astrophysical effects may introduce errors in our measurements of star-planet systems, which could ultimately impact planet formation and evolution theories. For example, \cite{shporer11} have shown that ignoring stellar surface convection in transit observations of the Rossiter-McLaughlin (RM) effect \citep{rossiter, mclaughlin, winn07} can lead to a deviation in the RVs on the m~s$^{-1}$ level, which the authors postulate will affect the measured spin-orbit alignment angle. Convection on the surface of solar-type stars results in a net convective blueshift (CB) of the spectral lines due to the fact that the uprising (blueshifted) granules are brighter and cover a greater surface area than the downflowing (redshifted) intergranular lanes (for the Sun this value is $\sim$ -300 m~s$^{-1}$; \citealt{dravins87a}). \cite{shporer11} produced a simple numerical model to illustrate this effect, wherein they considered the CB to be a constant value that varied across the stellar disc due to limb darkening and projected area. However, they acknowledged that such a model neglected effects from meridional flows, differential rotation, differences in CB for various stellar lines, as well as the dependence of the local observed CB on the centre-to-limb angle, $\theta$ (often denoted as $\mu = \cos(\theta$)), and hence may underestimate the total error in RM observations.  

\begin{center}
\begin{figure}[t!]
\centering
\includegraphics[trim=0.7cm 0.25cm 0.25cm 0.5cm, clip, scale=0.51]{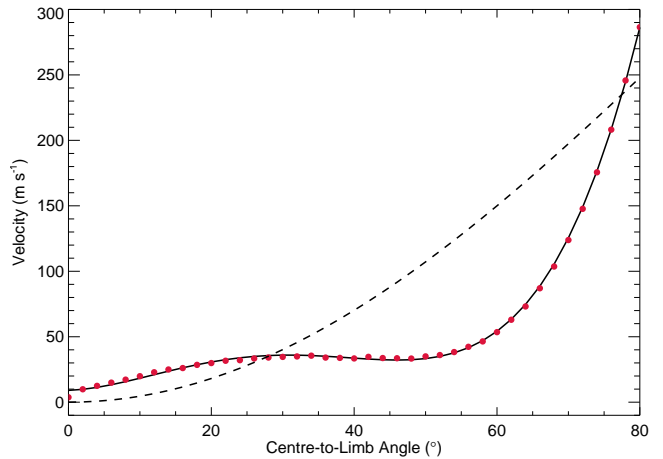}
\caption[]{The average granulation RVs, relative to disc centre, over an $\sim$ 80 minute time-series from the MHD solar simulation presented in \cite{cegla14b} as a function of stellar centre-to-limb angle (red dots). A solid (black) line illustrates a fourth order polynomial fit to the data, and a dashed (black) line illustrates the predicted variation in convective blueshift due solely to projected area for the Sun (i.e. a constant blueshift $\times \cos(\theta)$).} 
\label{fig:rvgrancorr}
\end{figure}
\end{center}

Indeed, solar observations and state-of-the-art 3D magnetohydrodynamic (MHD) simulations (coupled with radiative transport) clearly demonstrate that the observed variation in local CB may vary considerably from that predicted by projection effects alone (see Figure~\ref{fig:rvgrancorr} -- further discussed in Section~\ref{sec:mod}). This deviation is due to the corrugated nature of granulation. Across the stellar limb different aspects of the granulation are visible to the observer, e.g. when granulation is viewed near the stellar limb the tops of the granules and bottom of the intergranular lanes become hidden while the granular walls become visible. Hence, there are variations in the line-of-sight (LOS) velocities and flux that alter both the line shape and centroid, and result in RV variations in the observed local line profiles. 

In this paper, we use the centre-to-limb variation in CB predicted by a 3D MHD solar simulation, shown in Figure~\ref{fig:rvgrancorr}, to advance upon the analysis by \cite{shporer11}. We create stellar surface models that include not only stellar rotation and limb darkening, but also the variation in CB due to granulation corrugation (whilst accounting for the projected area at a given $\mu$). We inject a transiting planet into these stellar models and use the planet as a probe to resolve the CB variation in simulated Sun-as-a-star observations; this allows us to quantify the impact of ignoring the CB variation on RM measurements for Sun-like stars. We also independently quantify the error on the projected spin-orbit misalignment angle using the software tool SOAP-T \citep{oshagh13} as well as the Sun-as-a-star model code developed in \cite{cegla14b}. 

In Section~\ref{sec:mod}, we describe the two stellar models used throughout this paper. We present the RM waveform expected solely from a centre-to-limb variation in net CB for a Sun-like star in Section~\ref{sec:RMCB}. In Sections~\ref{sec:RMresid} and \ref{sec:spinorbit}, we quantify the deviation of the RM curve due to CB and the corresponding impact on the projected spin-orbit alignment angle. Finally, we conclude in Section~\ref{sec:conc}.

\vspace{-25pt}
\section{The stellar models}
\label{sec:mod}
Throughout this paper we use two stellar models, as each has one particular advantages over the other. In the first instance, we create a stellar grid following that used in \cite{cegla14b}, hereafter C14, while in the second instance we use the already established software tool SOAP-T. One advantage of the C14 model is that we can inject asymmetric line profiles to represent the stellar photosphere (as opposed to the strictly Gaussian profiles presently accepted by SOAP-T). Another advantage of the C14 model is that, in a forthcoming paper, we can include the variability of the ratio between granular and intergranular lanes on the stellar surface (as the granules evolve this ratio constantly changes and contributes a disc integrated RV variability on the order of 10s of cm~s$^{-1}$). On the other hand, the advantage of SOAP-T is that it is a well-tested numerical model currently used in the literature, and represents a typical numerical approach to modelling the RM waveform.

The C14 stellar grid was designed to incorporate line profiles from 3D MHD simulations. As such, a 3D sphere is covered in tiles with an area as close as possible to the area of the simulation snapshots; the 3D grid is then projected onto a 2D plane (as seen by the observer). The SOAP-T stellar grid, however, is constructed directly in the 2D plane, with a tile size optimised for planet transit analysis. Both codes inject into each tile a line profile (representative of the stellar photosphere) including the effects of limb darkening, projected area, and stellar rotational velocity shifts\footnote{For this work solid body rotation is assumed in order to isolate the impact from convection.}. A planetary transit is simulated by masking the tiles that correspond to the region behind the planet, and integrating over the stellar disc.

The main difference between these two models is that the C14 grid is tiled on a 3D surface and projected onto a 2D plane, whereas the SOAP-T grid originates in the 2D plane. This means that the C14 grid has a greater number of visible tiles near the stellar limb than it does near disc centre, whereas the SOAP-T grid has an even number of tiles throughout the stellar disc. Hence, some differences in the RM curves between the two models are expected since the tiling is slightly different. When we examined the residuals between the two stellar models, we concluded that although there were differences on the cm~s$^{-1}$ level that such differences were unlikely to affect the conclusions; see Appendix~\ref{appen:compare} for details.

In this paper, we only consider the impact of the local CB, without temporal variations. In the first instance, we modelled the local intrinsic line profiles as Gaussians. We use a quadratic limb-darkening law, where the coefficients (c$_1$ = 0.29, c$_2$ = 0.34) were determined by fitting the intensities from the MHD simulations in C14 (a quadratic limb-darkening law was chosen to match SOAP-T). The RVs for each observation were determined by the mean of a Gaussian fit to the disc-integrated line profiles. This technique was chosen as it is the same procedure used by the HARPS pipeline. Note that the HARPS pipeline operates on the CCF (cross-correlation function) created by the cross-correlation of the observed spectral absorption lines with a weighted template mask, and our disc-integrated profiles serve as a proxy for the CCFs. It is also important to note that a Gaussian fit only provides the true velocity centroid if the observed line profiles (and CCFs) are symmetric \citep[see][and Section~\ref{subsec:vsinifwhm} for more details]{cameron10}. Finally, each model was assigned the same star-planet properties; these are summarised in Table~\ref{tab:param}. In this work we modelled the transit of a 4 d hot Jupiter around a Sun-like star with an orbit that is aligned with the stellar spin axis. If not otherwise stated, the orbital inclination was 90$\degree$ (impact factor b = 0); this inclination was chosen so that the planet transited the maximum centre-to-limb positions across the stellar disc (note we do not suffer a degeneracy between the projected obliquity and the $v \sin i$, despite a zero impact factor, because we know the true stellar rotation of our model stars).

\begin{table}[h!]
\caption[]{Star and planet parameters in the model RM observations}
\centering
\begin{tabular}{c|c|c}
    \hline
    \hline
     Parameter & Star & Planet  \\
    \hline  
    Period  & variable\footnote{Stellar rotation was varied throughout, corresponding to $v \sin i = 1-10$~km~s$^{-1}$.} & 4 d\\
    Mass & 1 M$_{\odot}$ & 1 M$_{\rm J}$ \\
    Radius & 1 R$_{\odot}$ & 1 R$_{\rm J}$\\
    Eccentricity & -- & 0~~ \\
    Inclination & 90$\degree$ & -- \\
    Impact Factor & -- & variable\footnote{Initially b = 0, but in later sections it was varied to 0.25 and 0.5.} \\
    T$_{peri}$ & -- & 0~~ \\
    $\Omega$ & -- & 90$\degree$ \\
    $\gamma$ & -- & 0$\degree$ \\
   \hline

  \end{tabular}
\label{tab:param}
\end{table}
 
For each model we produced two sets of 93 observations, one with and one without the CB variation. These were centred about mid-transit with a cadence of 200~s (this gives close to 1~hr of out-of-transit time on either side of the transit). In the zero CB models, the intrinsic line profiles were only Doppler shifted by the appropriate stellar rotational velocity (no other line shifting mechanisms are included). For models with CB, the intrinsic profiles were shifted by both the stellar rotation and the simulated local CB variation from the solar simulations in C14.

The solar simulations in C14 were created with the MURaM code \citep{vogler05}, which has a simulation box corresponding to a physical size of 12 $\times$ 12 Mm$^2$ in the horizontal directions and 1.4 Mm in the vertical direction. The initial magnetic field was 200~G, which is only slightly higher than the unsigned average magnetic field in the `quiet' solar photosphere \cite[i.e. 130~G;][]{bueno04}. The photospheric plasma parameters from the MHD model were used to synthesise the 6302.5~$\AA$ Fe I line (with the STOPRO code). A time-sequence of 190 individual snapshots was produced, with a cadence of $\sim$ 30~s (except near the start of the simulation where the cadence was closer to 15~s). The sequence covers approximately 80 min., corresponding to $\sim$ 10-20 granular lifetimes. See \cite{cegla13} for further details on the simulation at disc centre. To create snapshots off disc centre, the horizontal layers of the simulation box were shifted to allow the LOS ray to penetrate the box from different angles. Centre-to-limb angles from 0-80$\degree$ were simulated in 2$\degree$ steps -- this step-size was largely set by computational constraints \citep{cegla15c}.

To determine the variation in local CB as a function of centre-to-limb angle, the line profiles from all snapshots in the time-sequence (at all stellar limb positions) were cross-correlated with one line profile from a single snapshot at disc centre. The disc centre template profile was chosen at random from the simulation time-series to set the zero-point for the cross-correlation, which was ultimately removed since we are only interested in the relative centre-to-limb variations. The peaks of the CCFs (from a second order polynomial fit) were used to determine the velocity shifts. To minimise the temporal influence (i.e. granulation evolution effects), all velocities at a given stellar limb position were averaged together over the 80 min. time-series\footnote{Note that shorter averaging timescales introduce scatter about the mean values over the entire (80-min.) time series, i.e. scatter about the red points plotted in Figure~\ref{fig:rvgrancorr}. For example, 5 minute averages introduce scatter of $\sim \pm$ 50 m s$^{-1}$ for positions $<$ 60$\degree$ and $\sim \pm$ 10 m s$^{-1}$ (or less) further toward the limb.}; the results are shown as red dots in Figure~\ref{fig:rvgrancorr}. To incorporate the CB variation in SOAP-T, we fit a fourth order polynomial to these points (solid line in Figure~\ref{fig:rvgrancorr}). For consistency, the same polynomial was used to introduce the CB velocity shifts in the C14 grid. Note we opted not to extrapolate the net CB beyond the 80$\degree$ centre-to-limb angle; this was because the slope of the polynomial fit at this limb angle is very steep (predicting an increase of 300~m~s$^{-1}$ from 80-90$\degree$) and since we do not know if this is truly physical we opted for a slight underestimation of the CB variation as opposed to a potentially large over estimation. All tiles with a centre-to-limb angle greater than 80$\degree$ were assigned the net CB corresponding to 80$\degree$. 

\section{RM waveform from centre-to-limb CB variations}
\label{sec:RMCB}
\begin{center}
\begin{figure}[t!]
\centering
\includegraphics[scale=0.45]{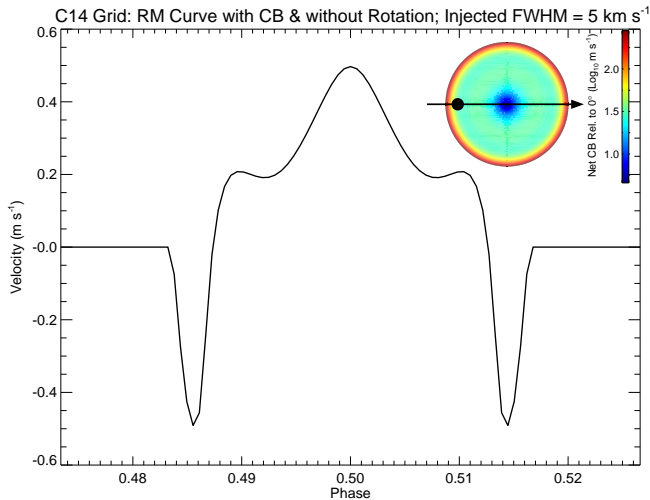}
\caption[]{Main: The measured RVs from a transit injected into the C14 grid for a non-rotating star (with Gaussian line profiles injected into the disc with a FWHM = 5~km~s$^{-1}$). Inset: Schematic of the planet transit across the stellar disc, colour-coded by the log of net convective velocities relative to disc centre.
\vspace{5pt}
} 
\label{fig:RM0}
\end{figure}
\end{center}

If the observed stellar surface velocities are only due to rotation, then a non-rotating star will have no RV anomaly during the planet transit and hence the RM waveform will be a flat line at zero velocity. However, in the presence of centre-to-limb CB variations, RV anomalies will still be apparent. To investigate the nature of such a signal, we injected the transiting planet into a system with the position-dependent net CB (shown in Figure~\ref{fig:rvgrancorr}) for a non-rotating star. Since SOAP-T is not designed to handle zero stellar rotation, this test was only performed using the C14 grid. In this instance, we injected Gaussian line profiles with a FWHM of 5~km~s$^{-1}$; this width was chosen as it is similar to the aforementioned 6302.5~$\AA$ Fe I line profile (from the 3D MHD solar simulations) at disc centre and therefore represents a realistic FWHM given the injected CB. The measured RVs for this set of observations is shown in Figure~\ref{fig:RM0} (alongside a schematic of the planet transit, colour-coded by the net convective velocities relative to disc centre). The RVs near ingress and egress are blueshifted since the planet obscures the local CBs with the highest redshifts (relative to disc centre) and redshifts near mid-transit where the planet obscures more blueshifted regions of the stellar disc. Hence, from Figure~\ref{fig:RM0} we can see that a local variation in CB contributes to the RV anomaly observed during transit, and leads to a non-zero RM waveform even when no stellar rotation is observed (the exact shape and amplitude of this waveform will depend on the planet-to-star ratio and the convective properties of the star).

\begin{center}
\begin{figure*}[t!]
\begin{center}
\includegraphics[scale=0.44]{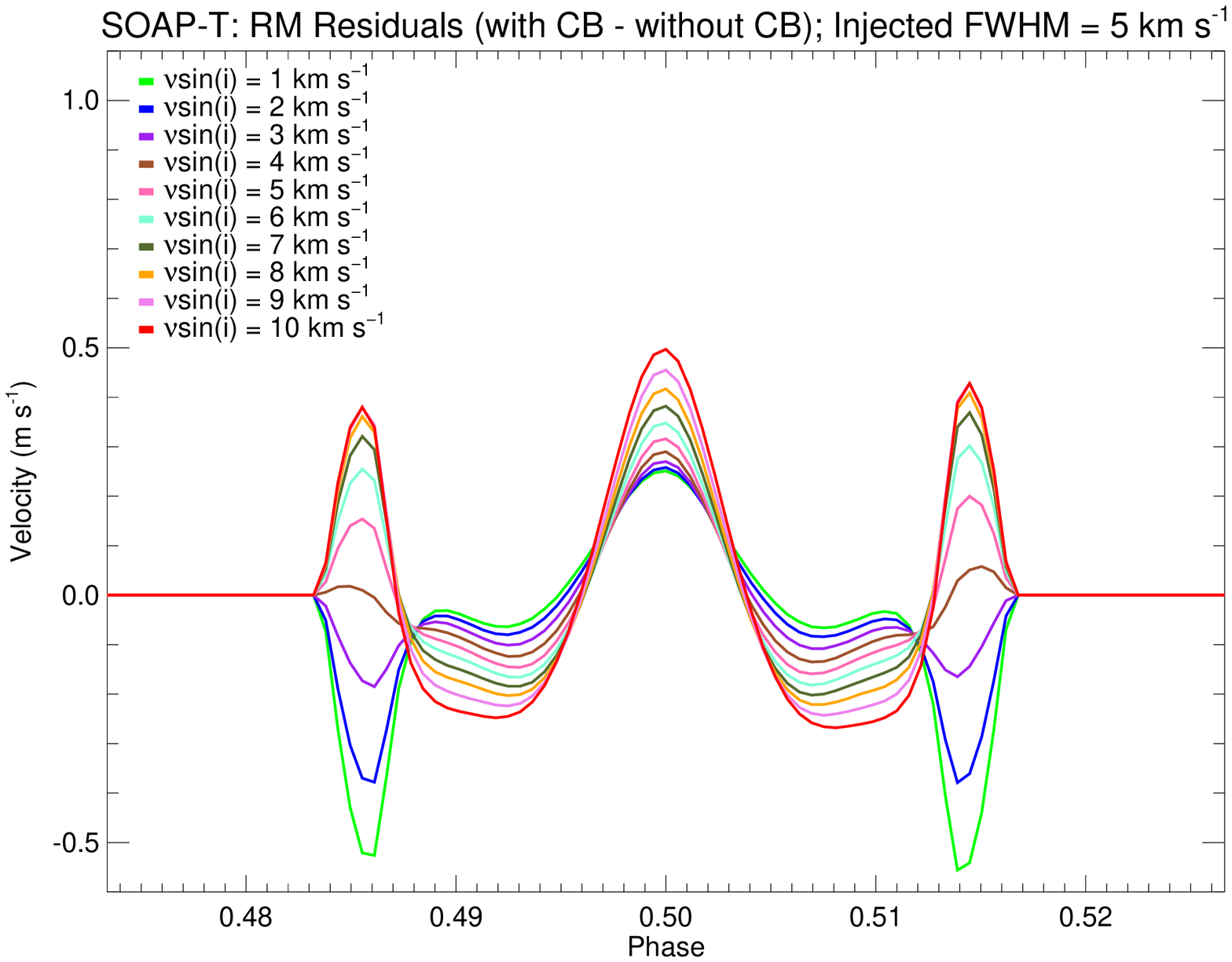}
\includegraphics[scale=0.44]{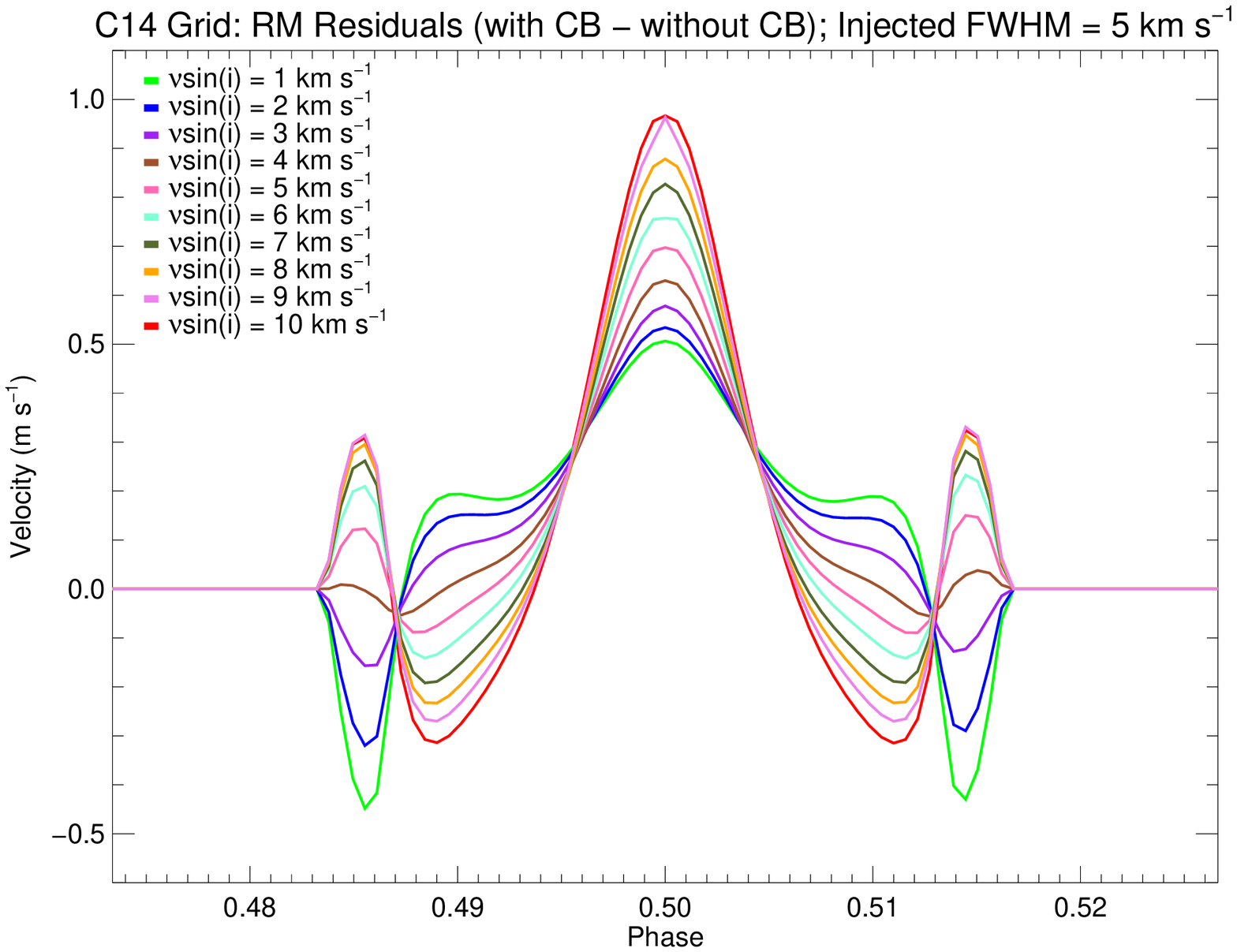}
\caption{Residual RM curves for both SOAP-T (left) and the C14 grid (right), where the residuals are defined as observations with CB - observations without CB. In both cases the FWHM of the injected profile is 5~km~s$^{-1}$ and the $v \sin i$ is varied from 1 - 10 km~s$^{-1}$. 
} 
\label{fig:RMresid}
\end{center}
\end{figure*}
\end{center}
\vspace{-30pt}

It is also important to note that the inclusion of the CB variation across the stellar limb causes an asymmetry in the disc-integrated line profiles. This asymmetry is seen even for out-of-transit observations and even if the intrinsic profiles are Gaussian. Moreover, it leads to non-zero out-of-transit RVs in the models with CB (that are removed as we are only interested in the relative RVs). This effect is similar to the `C'-shape bisector seen in stellar observations of cool stars \citep{gray05}. In this instance, the asymmetry arises from the combination of limb-darkening and radial CB variation, i.e. the brightest regions of the disc (near the centre) will have profiles with a much bluer net CB compared to the darker regions of the disc (near the limb), that will have profiles with a local CB that is redshifted relative to the value at disc centre. Hence, integrated annuli near disc centre will have a different brightness and net RV shift compared to those near the limb, and summing over these annuli creates the asymmetry. The level of asymmetry will vary based on the FWHM of the injected line profile and the stellar rotation. This asymmetry also depends on the shape and amplitude of the centre-to-limb CB, which is expected to increase with decreasing magnetic field (as the convective flows will flow more freely), and on the observed stellar lines and the spectral type (note varying these parameters is beyond the scope of this paper). 

\section{RM curves with and without CB effects}
\label{sec:RMresid}

\subsection{The impact of $v \sin i$ and intrinstic profile FWHM}
\label{subsec:vsinifwhm}
The observed RVs depend not only on the given star-planet system (i.e. star/planet masses, radii, orbital separation, inclination, and alignment), but also on the line broadening inherent to the star as this impacts the observed line profile asymmetries, and hence the measured line centre. The disc-integrated profile width/shape depends on the observed stellar rotation (i.e. $ v \sin i$) and the intrinsic profile width (set largely by convective broadening, i.e. `macroturbulence', and thermal broadening -- and to a lesser extent a number of collisional broadening mechanisms), as well as the instrumental profile. Consequently, we explored the residuals between observations with and without CB (i.e. $RM_{withCB} - RM_{withoutCB}$) for systems with a variety of stellar rotation rates and injected profile FWHMs. We remind the reader that at this stage all models are injected with local Gaussian profiles (though the disc-integrated CB model profiles are asymmetric).

The residual RM curves for stars with a fixed intrinsic profile FWHM of 5~km~s$^{-1}$ and $v \sin i$ from 1 - 10 km~s$^{-1}$ are shown in Figure~\ref{fig:RMresid} for both stellar models (left: SOAP-T; right: C14 grid).  One might expect the amplitude of these residuals to decrease once the LOS stellar rotation is large enough to dominate the RVs over the variation in local CB. Interestingly, this is not observed (however, do note that this is the case if the residuals are normalised by the maximum amplitude of the RM signal). The amplitude of these residuals varies from $\sim$0.1-1~m~s$^{-1}$, depending on $v \sin i$, which will be important for, and detectable with, future instruments such as ESPRESSO.\footnote{We note that in this RV regime, other physical effects such as gravitational microlensing of the transiting planet may also need to be taken into account \citep{oshagh14}.} For the slowly rotating stars, these residuals show a similar overall behaviour to that seen in Figure~\ref{fig:RM0}. However, as $v \sin i$ becomes larger than the injected profile FWHM the ingress and egress regions switch from blueshifted to redshifted. The origin for this unexpected behaviour is not clear, but could be related to the errors introduced when fitting a Gaussian function to an asymmetric profile and/or because the limb contribution (where the net CB is most redshifted) impacts the disc-integrated profile more once the $v \sin i$ is greater than the intrinsic broadening \citep{gray85, smith87, dravins90b, bruning90}. Greater stellar rotation also leads to an increased redshift at mid-transit and a decreased redshift in the regions between ingress/egress and mid-transit. Hence, a larger stellar rotation increases the overall amplitude between the local maxima and minima in this region (which excludes the ingress/egress points). The behaviour of these residuals is similar in both SOAP-T and the C14 grid, though the exact shape and amplitude of the curves does differ slightly (likely due to the tiling differences). We also found a very similar, though opposite, behaviour in the residuals when we held the $v \sin i$ constant (at 5~km~s$^{-1}$) and varied the injected line profile FWHM; this is because the shape of the disc-integrated profile depends heavily on both the rotational broadening and the width of the intrinsic profiles on the stellar surface.

Note that unlike the RM curve in Figure~\ref{fig:RM0} (which had CB variation, but no stellar rotation), these residuals are not symmetric about mid-transit (in agreement with that found in \citealt{dravins15}); this is particularly evident in the ingress/egress regions. From a purely mathematical point-of-view, these residuals should be symmetric as they are the result of an odd function (stellar rotation RVs) being subtracted from a function which is the sum of an odd and even function (stellar rotation RVs + radial CB variations). To understand the non-symmetric residuals, it is important to keep in mind that the RVs are measured by fitting a Gaussian function to the observed disc-integrated line profile. 

Fitting a Gaussian function to an asymmetric line profile does not provide the true velocity centroid of the visible light. If we are interested in relative velocity changes then this offset does not matter, as long as the asymmetry remains the same. For a (model) star with CB and without stellar rotation (see Section~\ref{sec:RMCB}), the asymmetries in the disc-integrated line profiles will change during transit. However, since the CB is an even function, these asymmetries will be the same for a given centre-to-limb position, and will lead to symmetric RVs (for aligned star-planet systems) as the offsets in the true velocity centroid will also be symmetric. For (model) stars with stellar rotation and without CB, the asymmetries will be mirror images of one another about mid-transit (hence the typical RM effect) and will lead to RVs that are symmetric about mid-transit\footnote{Note that although these RVs will be symmetric about mid-transit, the errrors introduced from the Gaussian fit can still bias the analysis. For example, \cite{triaud09} proposed that the errors introduced by the Gaussian approximation were responsible for the m~s$^{-1}$ residuals between their measured RVs and RM model for the transit of HD 189733 b. Additionally, they argued that if these errors were not taken into account the measured $v \sin i$ could be off by as much as $\sim$ 300~m~s$^{-1}$ for this system.}. For stars with both CB and stellar rotation, the asymmetries are not the same for a given centre-to-limb angle, nor are they mirror images of one another. As a result, the offset in absolute velocity as measured by the Gaussian function will vary in a complex way. Hence, the RVs will not represent perfectly the sum of an odd and even function and therefore the RM residuals between the observations with and without CB will not be perfectly symmetric (however, note that the asymmetry in the residuals found here is on the $<$10~cm~s$^{-1}$ level). This is a fundamental limitation of the Gaussian fit RV technique, which can introduce offsets/systematic errors into high precision RV studies \citep[e.g. see][and references therein]{triaud09,cameron10, miller10}. Furthermore, both \cite{hirano10} and \cite{boue13} have shown that the errors introduced by the Gaussian approximation will scale with both $v \sin i$ and intrinsic profile width. Accordingly, we believe that the increase in amplitude of the residuals in Figure~\ref{fig:RMresid} is at least partially due to the errors introduced by fitting Gaussians to the disc-integrated profiles in order to obtain the RV. 

\begin{center}
\begin{figure}[b!]
\begin{center}
\includegraphics[scale=0.44]{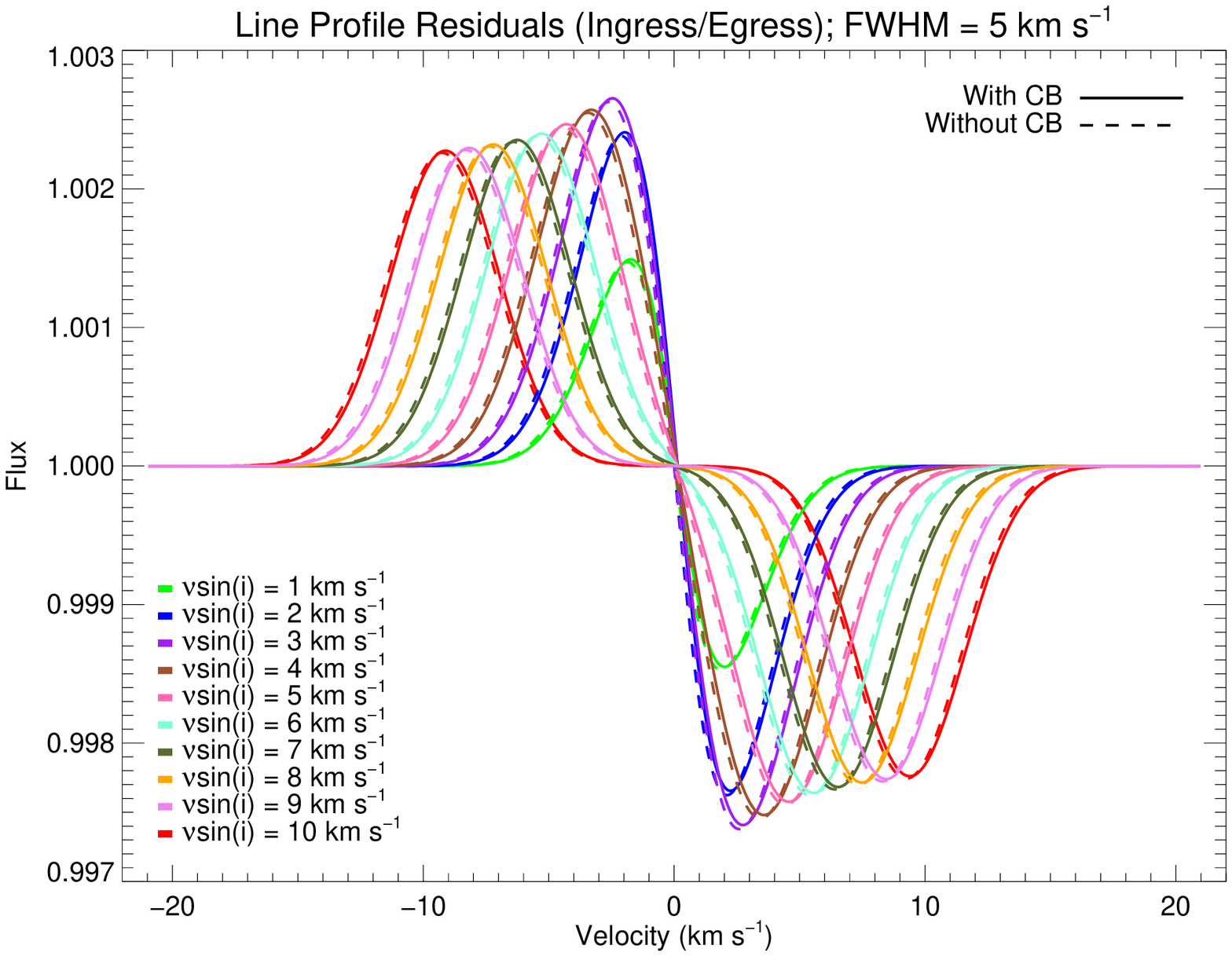}
\includegraphics[scale=0.44]{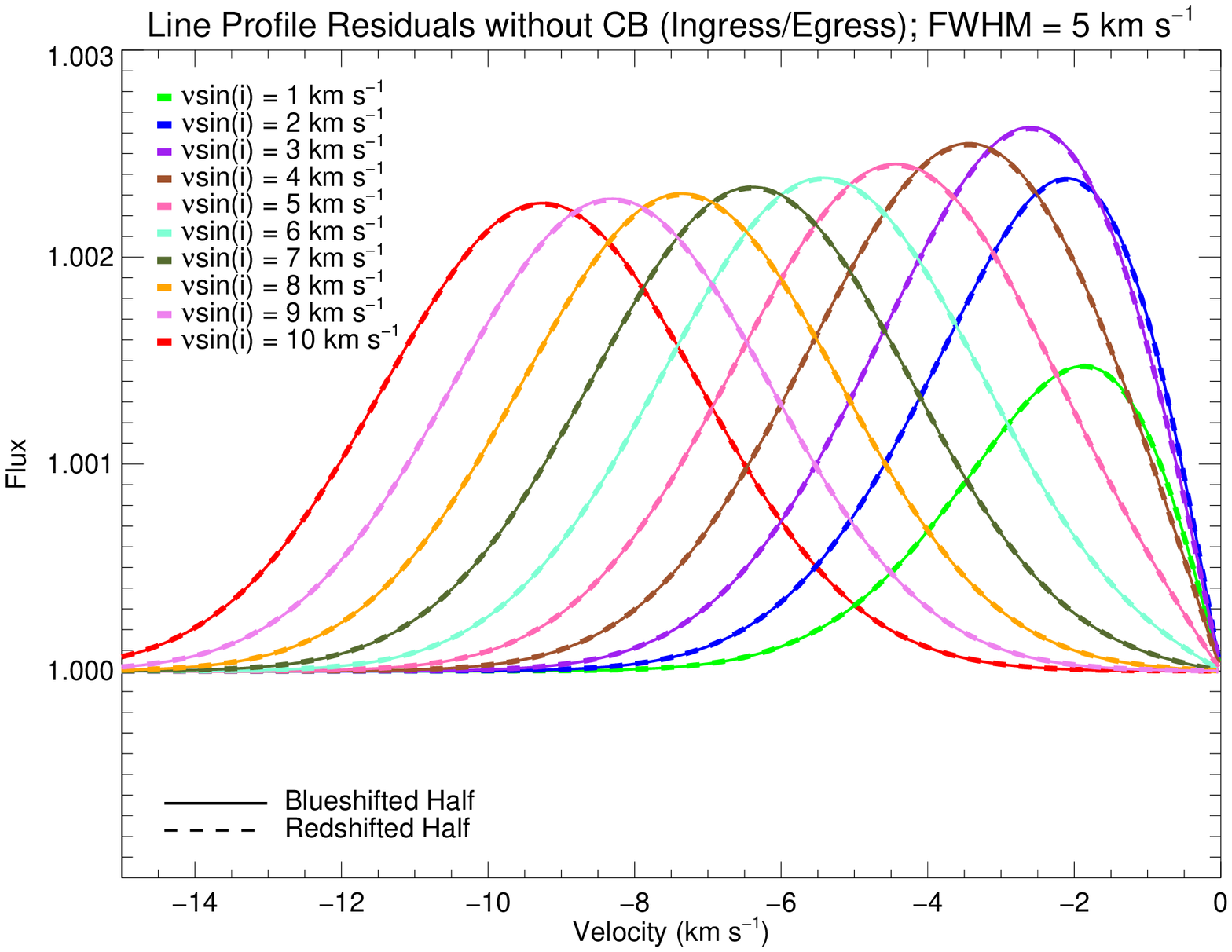}
\includegraphics[scale=0.44]{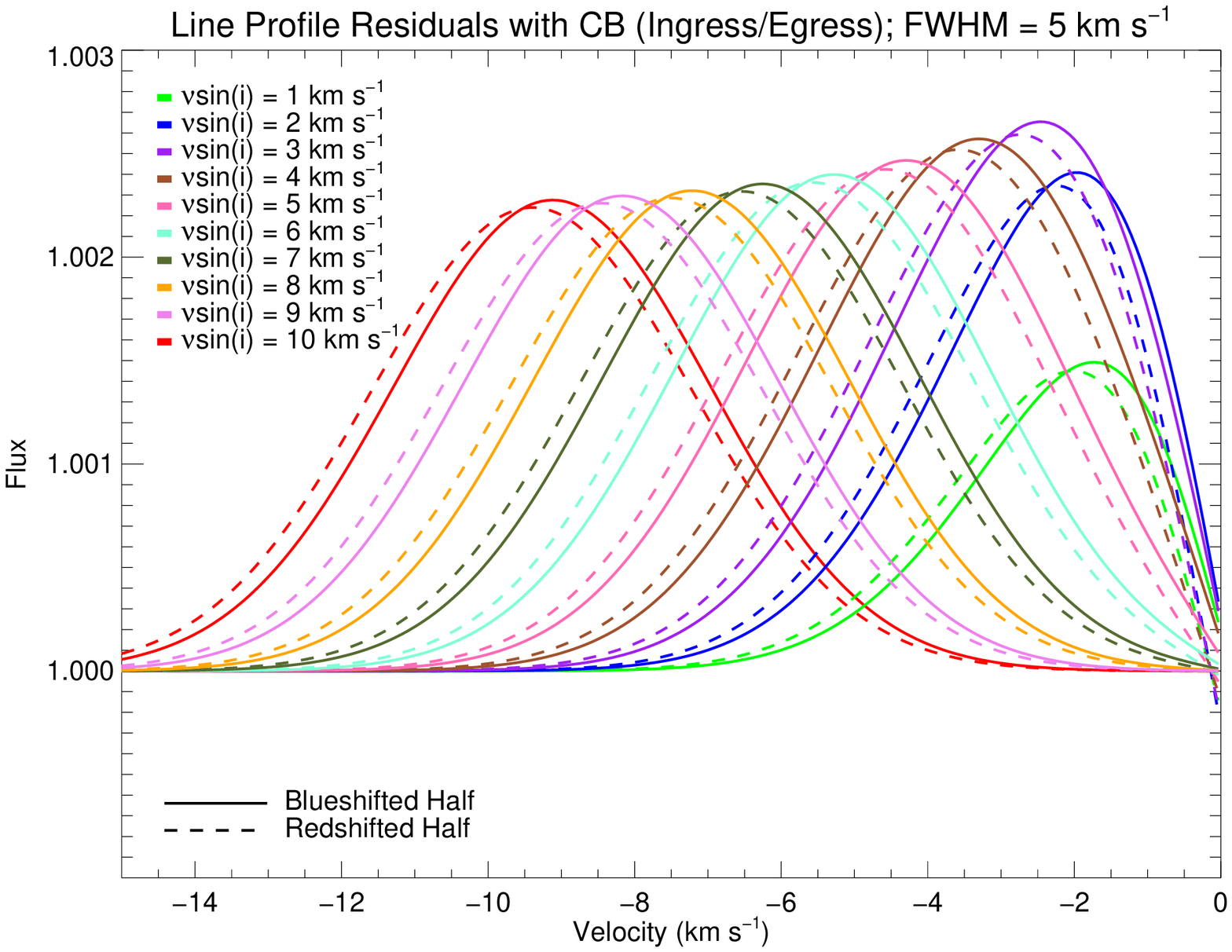}
\caption{Top: Residuals from a line profile at ingress divided by the equivalent profile at egress for observations with (solid) and without (dashed) CB for stars with varying $v \sin i$. Middle: Same as top, but only for model without CB and where the redshifted flux values have been flipped, reversed and over plotted as dashed lines. Bottom: Same as middle, but for the model with CB included.} 
\label{fig:prflresid}
\end{center}
\end{figure}
\end{center} 
\vspace{-10pt}

To illustrate the shapes of the disc-integrated line profiles on either side of mid-transit, we show profiles at ingress divided by those at egress (where the asymmetry in the RM residuals seen in Figure~\ref{fig:RMresid} is largest) in the top plot in Figure~\ref{fig:prflresid}; the two bumps are due to the Doppler shift between the profiles and any difference in line shape. These are presented only for the C14 grid and only for the observations which varied the $v \sin i$; analysis using SOAP-T (and varying the FWHM) showed similar results. If the above reasoning is correct (and the Gaussian approximation is responsible for the asymmetry in the RM residuals) then the models that include both rotation and CB must create profiles that differ in a way that is not a simple mirror image. If the profiles are mirror images of one another, then flipping and reversing all the flux values that correspond to the redshifted velocity space should result in points that lie exactly on top of those in the blueshifted velocity space. This test is shown in the middle and bottom plots in Figure~\ref{fig:prflresid} for the models excluding and including CB, respectively. From these we see that the profiles without CB are in fact mirror images of one another (as expected from stellar rotation alone). We also see that the profiles including CB shifts are definitely not mirror images of one another. Hence, this allows for the possibility that the errors in the RV measurement due to the Gaussian fit may differ between these profiles and could therefore produce RV shifts that are not equal in magnitude.

\begin{center}
\begin{figure*}[t!]
\begin{center}
\includegraphics[scale=0.44]{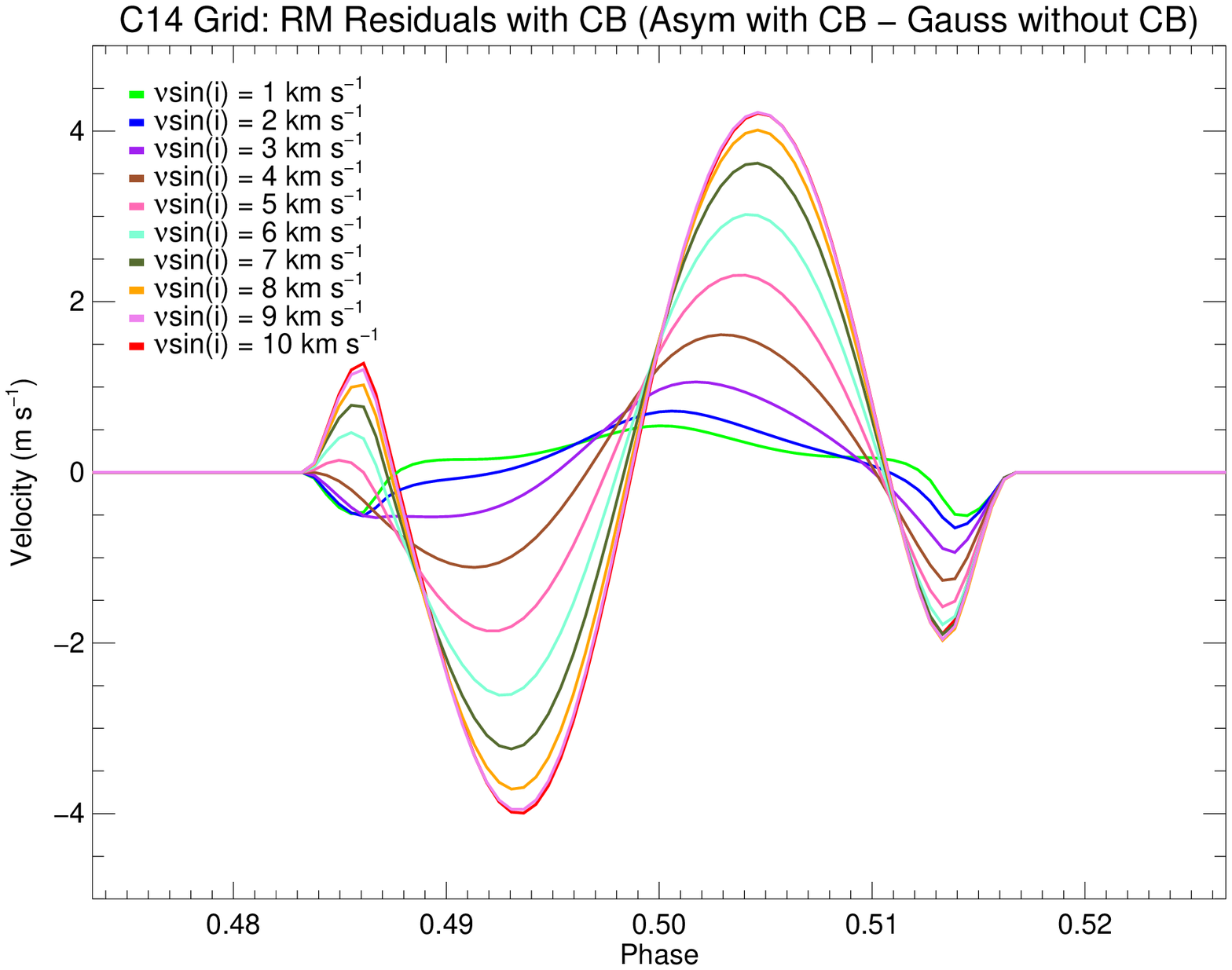}
\includegraphics[scale=0.44]{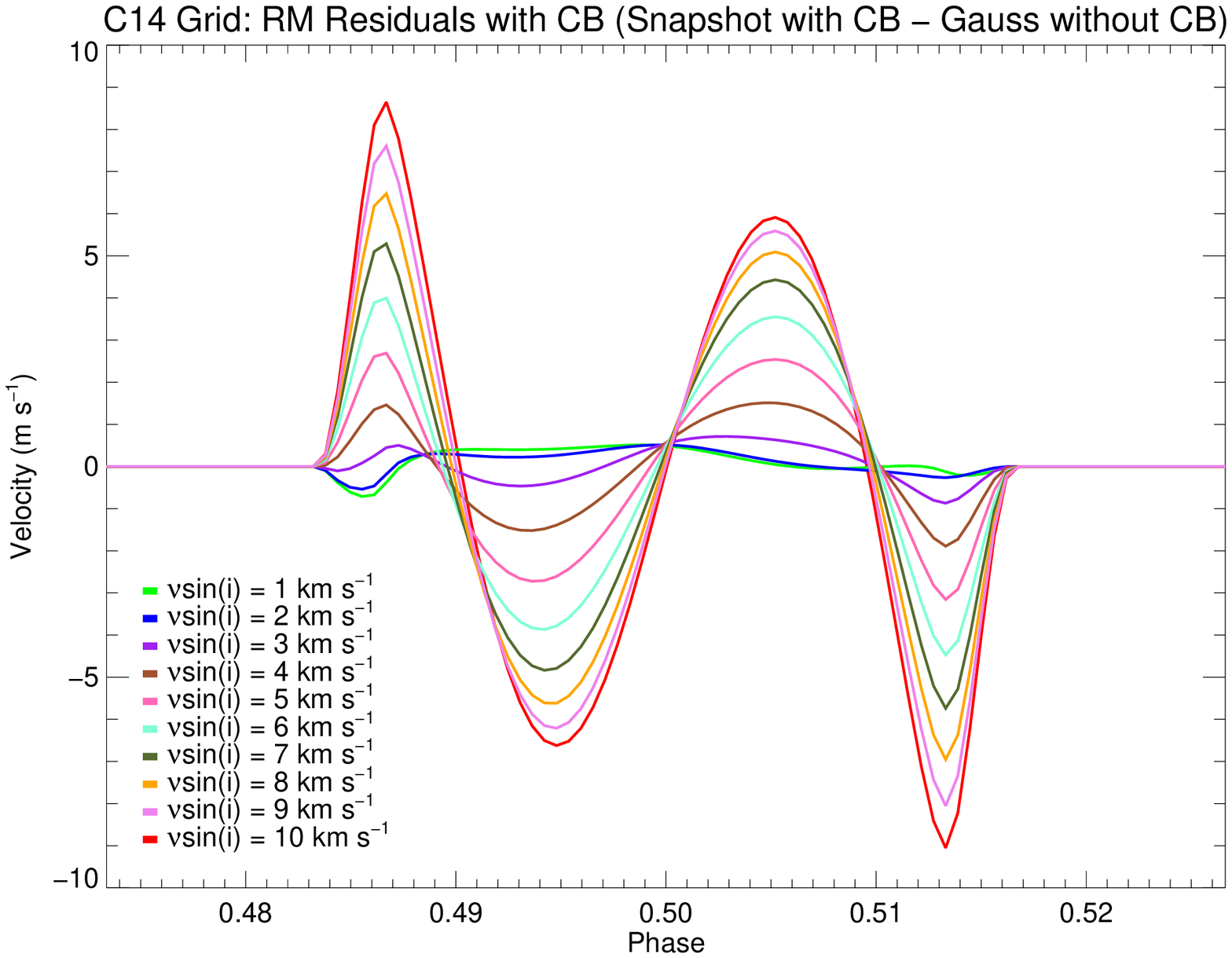}
\caption{Left: Residual RM curves for model observations where $v \sin i$ was varied from 1 - 10 km~s$^{-1}$. The residuals were constructed as the difference between the model stars where the grid was injected with Gaussian line profiles with a FWHM of 5~km~s$^{-1}$ excluding CB, and model stars injected with one asymmetric line profile chosen at random from a disc centre snapshot of the radiative 3D MHD solar simulation including CB variations. Right: Same as Left, but injected with asymmetric profiles chosen from the (same) solar simulation snapshot when inclined from 0-80$\degree$ on the stellar disc (rather than injecting  the disc centre profile everywhere).} 
\label{fig:RMresidwasym}
\end{center}
\end{figure*}
\end{center}
\vspace{-30pt}

\subsection{The impact of line profile shape/symmetry}
\label{subsec:asymprfl}
We also explored the impact of injecting an asymmetric intrinsic line profile into the stellar disc. In the first case, we injected one line profile throughout the C14 grid randomly chosen from a disc centre snapshot in the solar (MHD) simulation time-series (SOAP-T does not yet have the ability to accept asymmetric line profiles); such a profile was chosen as the asymmetries are realistic and representative of those produced by solar surface (magneto-)convection. The left plot in Figure~\ref{fig:RMresidwasym} shows the residuals of the RM curve comparing model stars with Gaussian intrinsic line profiles and no CB, to those that include CB and have asymmetric line profiles for different $v \sin i$ (1 - 10 km~s$^{-1}$). For very slow rotators (i.e. $v \sin i \le $ 2 km~s$^{-1}$) the differences are $ \le \sim$ 0.5 m~s$^{-1}$ and therefore difficult to detect with current instrumentation (but not beyond the reach of future spectrographs). For faster rotators (3 km~s$^{-1}$ $ \le v \sin i \le $ 10 km~s$^{-1}$) the differences can be as large as $\sim$ 1 - 4 m~s$^{-1}$ in amplitude, and are therefore readily detectable with current spectrographs such as HARPS, HARPS-N, and HIRES (note, that for higher $v \sin i$ the total amplitude of the signal will also be higher, and the relative impact on the RM modelling may be less significant). For the faster rotators, the effects of the CB variation become less visible and the impact of the profile shape dominates. However, what might be more significant than the amplitude of the residuals is the asymmetries present between the RVs on either side of mid-transit, as well as the net velocity at mid-transit (which is non-zero and dependent on the $v \sin i$). These asymmetries in the RM curve may be incorrectly interpreted as a non-zero spin-orbit misalignment since the impact factor is usually fixed by the light curve. 

In reality, the intrinsic line shape changes as a function of centre-to-limb angle. To examine this effect, we used the same simulation snapshot as before (to avoid changes from granular evolution), but injected profiles from the snapshot when inclined as close as possible to the same centre-to-limb angles as the tiles in the stellar grid. The residuals between the stars that excluded CB and had intrinsic Gaussian profiles and those stars with CB and the limb-dependent asymmetric profiles are shown in the right of Figure~\ref{fig:RMresidwasym}. These residuals are much larger in amplitude than any of the previous, with RVs near 10~m~s$^{-1}$ for the fastest rotators. If the observed CCF of the local stellar photosphere varies as much as the injected line profile from the radiative 3D MHD simulation, then these differences should be easily detectable (note that an observed CCF may experience less centre-to-limb variability since it is created from the information content of thousands of lines that have a variety of granulation sensitivity). We note that a high sampling rate at ingress/egress would be beneficial for such an empirical verification since these regions experience the largest discrepancies. 

\section{The impact of centre-to-limb CB variations on spin-orbit misalignment measurements}
\label{sec:spinorbit}
In the previous section we have shown that ignoring the effects of CB and the formation of asymmetric line profiles can alter predicted RVs by 10s of cm s$^{-1}$ to m s$^{-1}$. However, the RM effect is primarily studied to determine the alignment of planetary systems with respect to the host star spin axis.

As such, we wish to quantify the impact of the convective centre-to-limb variation on measurements of the projected spin-orbit alignment angle, $\lambda$. To do so we simulated the aforementioned aligned($\lambda = 0\degree$) star-planet system with a stellar model that included the CB variation to act as our $observed$ data. To fit these simulated observations, we applied models that assumed no convective blueshift terms, and intrinsic Gaussian profiles -- inline with traditional RM studies. To fit the data, $\lambda$  was allowed to vary $\pm 30\degree$ in 1$\degree$ intervals; the fits were generated using both the C14 and SOAP-T packages\footnote{Note we did further test fits with 10$\degree$ steps in $\lambda$ from 40-90$\degree$ to ensure the fits did not change outside the chosen $\pm 30\degree$ fit interval.}.

Since the RM residuals between models with and without CB are dependent on the stellar rotation, we performed this comparison for both a slow ($v \sin i = 2$~km~s$^{-1}$) and moderately rapidly ($v \sin i = 6$ km s$^{-1}$) rotating star. The RM signal is also dependent on the correct modelling of the intrinsic profile shape, hence we repeated these tests while varying the intrinsic profile in the stellar model representing the $observed$ data. The injected intrinsic profiles were either Gaussian (matching the $fitted$ data), or a single asymmetric profile (from the MHD simulation at disc centre), or a range of asymmetric profiles (from a single MHD snapshot of granulation, inclined from 0-80$\degree$ on the stellar disc). 

\begin{table*}[t!]
\caption[]{Recovered obliquities of the aligned model RM observations as determined by $\chi^2$ minimisation}
\centering
\begin{tabular}{|c||c|c|c|}
\hline
\hline
Stellar Grid & \multicolumn{3}{ |c| }{C14} \\
\hline
Impact Factor & \multicolumn{1}{ c| }{b=0.0} & \multicolumn{1}{ c }{b=0.25} & \multicolumn{1}{ |c| }{b=0.5} \\

\hline
$v \sin i$& \multicolumn{3}{ |c| }{} \\ 
    \hline  
\multicolumn{4}{ |c| }{Intrinsic Profile Represented by a Gaussian}\\
    \hline
    2 km s$^{-1}$ & $\lambda =-5^{+17}_{-6}$$\degree$; $ \chi^2_{r}=1.19$ & $\lambda = 2^{+1}_{-2}$$\degree$; $\chi^2_r=1.06$ & $\lambda = 0.3^{+1.5}_{-0.9}$$\degree$; $ \chi^2_{r}=1.07$ \\
     6 km s$^{-1}$ & $\lambda =0^{+8}_{-7}$$\degree$; $ \chi^2_{r}=1.25$ & $\lambda = 0.5\pm 0.6$$\degree$; $\chi^2_r = 1.05$ & $\lambda =0.1^{+0.3}_{-0.2}$$\degree$; $ \chi^2_{r}=1.02$\\
   \hline
\multicolumn{4}{ |c| }{Intrinsic Profile Represented by a Single Asymmetric Profile}\\
   \hline
    2 km s$^{-1}$ & $\lambda =3^{+10}_{-16}$$\degree$; $ \chi^2_{r}=1.39$ & $\lambda = 2\pm 2$$\degree$; $\chi^2_r = 1.22$ & $\lambda =1\pm 1$$\degree$; $ \chi^2_{r}=1.19$\\
    6 km s$^{-1}$ & $\lambda =\pm 23 \pm 1$$\degree$; $ \chi^2_{r}=2.72$ & $\lambda = 1.5^{+0.7}_{-0.3}$$\degree$; $\chi^2_r = 7.53$ & $\lambda =0.7^{+0.2}_{-0.4}$$\degree$; $ \chi^2_{r}=6.56$\\
   \hline
\multicolumn{4}{ |c| }{Intrinsic Profile Represented by a Range of Asymmetric Profiles}\\
   \hline
    2 km s$^{-1}$ & $\lambda = 3^{+6}_{-12}$$\degree$; $ \chi^2_{r}=1.21$ & $\lambda = 1^{+2}_{-1}$$\degree$; $\chi^2_r = 1.16$ & $\lambda = 1 \pm 1$$\degree$; $ \chi^2_{r}=1.37$\\
    6 km s$^{-1}$ & $\lambda =\pm 27 \pm 1$$\degree$; $ \chi^2_{r}=4.41$ & $\lambda = 0.1^{+0.6}_{-0.5}$$\degree$; $\chi^2_r = 18.85$ & $\lambda =0.5^{+0.2}_{-0.4}$$\degree$; $ \chi^2_{r}=10.96$\\
\hline
\hline
Stellar Grid & \multicolumn{3}{ |c| }{SOAP-T} \\
    \hline  
\multicolumn{4}{ |c| }{Intrinsic Profile Represented by a Gaussian}\\
    \hline
    2 km s$^{-1}$ & $\lambda =-3^{+8}_{-5}$$\degree$; $ \chi^2_{r}=1.18$ & $\lambda = -2^{+10}_{-12}$$\degree$; $\chi^2_r=1.05$ & $\lambda =-1^{+8}_{-7}$$\degree$; $ \chi^2_{r}=1.18$ \\
     6 km s$^{-1}$ & $\lambda =-6^{+11}_{-6}$$\degree$; $ \chi^2_{r}=1.05$ & $\lambda = 0^{+4}_{-6}$$\degree$; $\chi^2_r = 1.97$ & $\lambda =0 \pm 3$$\degree$; $ \chi^2_{r}=1.14$\\
   \hline
\multicolumn{4}{ |c| }{Intrinsic Profile Represented by a Single Asymmetric Profile}\\
   \hline
    2 km s$^{-1}$ & $\lambda =-11^{+3}_{-2}$$\degree$; $ \chi^2_{r}=1.27$ & $\lambda = -1^{+12}_{-14}$$\degree$; $\chi^2_r = 1.32$ & $\lambda =-4^{+6}_{-8}$$\degree$; $ \chi^2_{r}=1.03$\\
    6 km s$^{-1}$ & $\lambda =-25,+23^{+12}_{-5} $$\degree$; $ \chi^2_{r}=2.79$ & $\lambda = -2^{+10}_{-8}$$\degree$; $\chi^2_r = 3.95$ & $\lambda =-1^{+6}_{-5}$$\degree$; $ \chi^2_{r}=2.33$\\
   \hline
\multicolumn{4}{ |c| }{Intrinsic Profile Represented by a Range of Asymmetric Profiles}\\
   \hline
    2 km s$^{-1}$ & $\lambda = -5^{+8}_{-7}$$\degree$; $ \chi^2_{r}=1.02$ & $\lambda = 5^{+8}_{-7}$$\degree$; $\chi^2_r = 1.02$ & $\lambda = -2^{+9}_{-8}$$\degree$; $ \chi^2_{r}=1.06$\\
    6 km s$^{-1}$ & $\lambda = \pm 27^{+10}_{-5}$$\degree$; $ \chi^2_{r}=2.75$ & $\lambda = 0^{+8}_{-7}$$\degree$; $\chi^2_r = 4.31$ & $\lambda =-1 \pm 6$$\degree$; $ \chi^2_{r}=3.63$\\
   \hline

    \end{tabular}
\label{tab:lamb}
\end{table*}

We decided to also test two non-zero impact factors. This is because for an impact factor of zero, the symmetry of the RM signal is unaffected by the spin-orbit alignment if one assumes the observed RV signal originates only from stellar rotation and the intrinsic profile is Gaussian. In this scenario, changing the alignment only alters the amplitude of the RM signal (similar to a change in stellar rotation rate -- note since we know the true stellar rotation we do not suffer the usual degeneracies between $v \sin i$ and the projected obliquity when fitting a system with b = 0). On the other hand, the shape of the RM signal from transits with non-zero impact factors are influenced by the spin-orbit alignment (and hence are typically targeted for RM observations). Including the CB variation (and asymmetric intrinsic profiles) alters the symmetry of the RM signal, regardless of the impact factor. Hence, for a more complete view of the influence of convection on the measurements of $\lambda$ we also consider impact factors of 0.25 and 0.5. Exploring additional impact factors is beyond the scope of this paper and will be pursued in forthcoming publications. 

To determine the impact on the measured $\lambda$, we performed a $\chi^2$ minimisation between the models with CB and those without. Before doing so, we added Gaussian noise at the 0.5 m s$^{-1}$ level (consistent with high-quality HARPS observations) to the models with CB acting as the $observed$ data. The $\chi^2$ calculation was then determined in a Monte-Carlo fashion by repeating the calculation 1000 times for different generations of random noise. The average $\chi^2$ of the 1000 generations was then used to compare the models with CB to the models without CB, with the best-fit model corresponding to the $\chi^2$ minimum. The obliquities that correspond to the best-fit models can be found in Table~\ref{tab:lamb}, alongside the reduced $\chi^2$ (shown to illustrate the goodness of fit between models, hereafter $\chi^2_r$). The error quoted on $\lambda$ corresponds to the 3$\sigma$ confidence interval on the $\chi^2$ minimum (i.e. since we have 1 free parameter, $\lambda$, this interval corresponds to a $\Delta \chi^2 = 9$); note that at times an uncertainty of 0$\degree$ arose due to the limitation of our 1$\degree$ step-size in $\lambda$ -- for these systems the fitted $\lambda$ was allowed to vary in finer 0.1$\degree$ steps.

If the true intrinsic profile can indeed be represented by a Gaussian function, then our best-fit models indicated little or no spin-orbit misalignment. This was regardless of the impact factor and $v \sin i$ chosen, with each scenario achieving a $\chi^2_r$ near 1 --  though there was one instance when comparing with SOAP-T that the $\chi^2_r$ is closer to 2 (fast rotator when b = 0.25). Additionally, for the C14 grid we found the 3$\sigma$ confidence interval corresponded to a variation in $\lambda$ of $\sim$10$\degree$  when b = 0, but decreased to a variation of only 1-2$\degree$ for non-zero impact factors. 

\begin{center}
\begin{figure*}[t!]
\begin{center}
\includegraphics[scale=0.44]{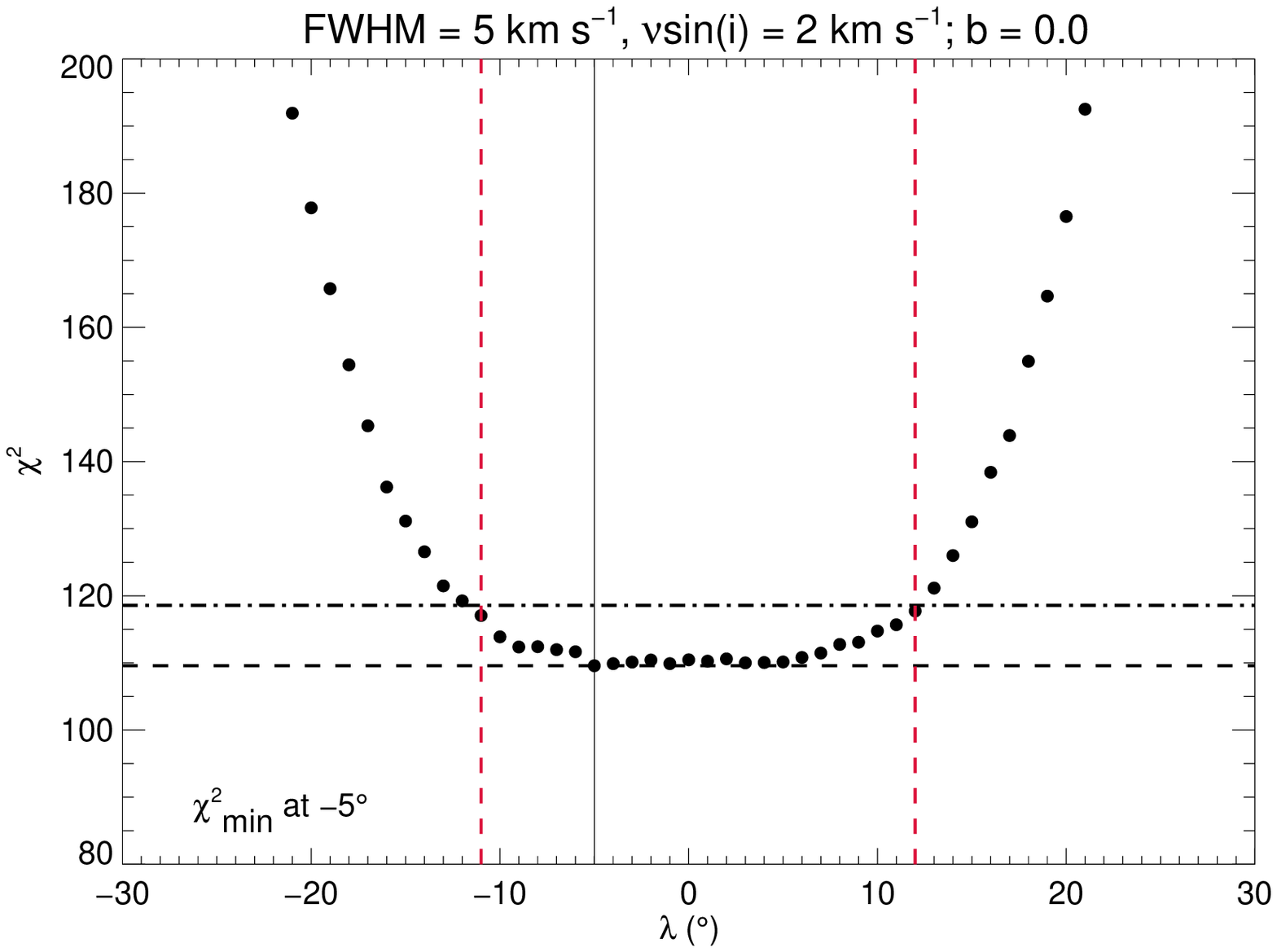}
\includegraphics[scale=0.44]{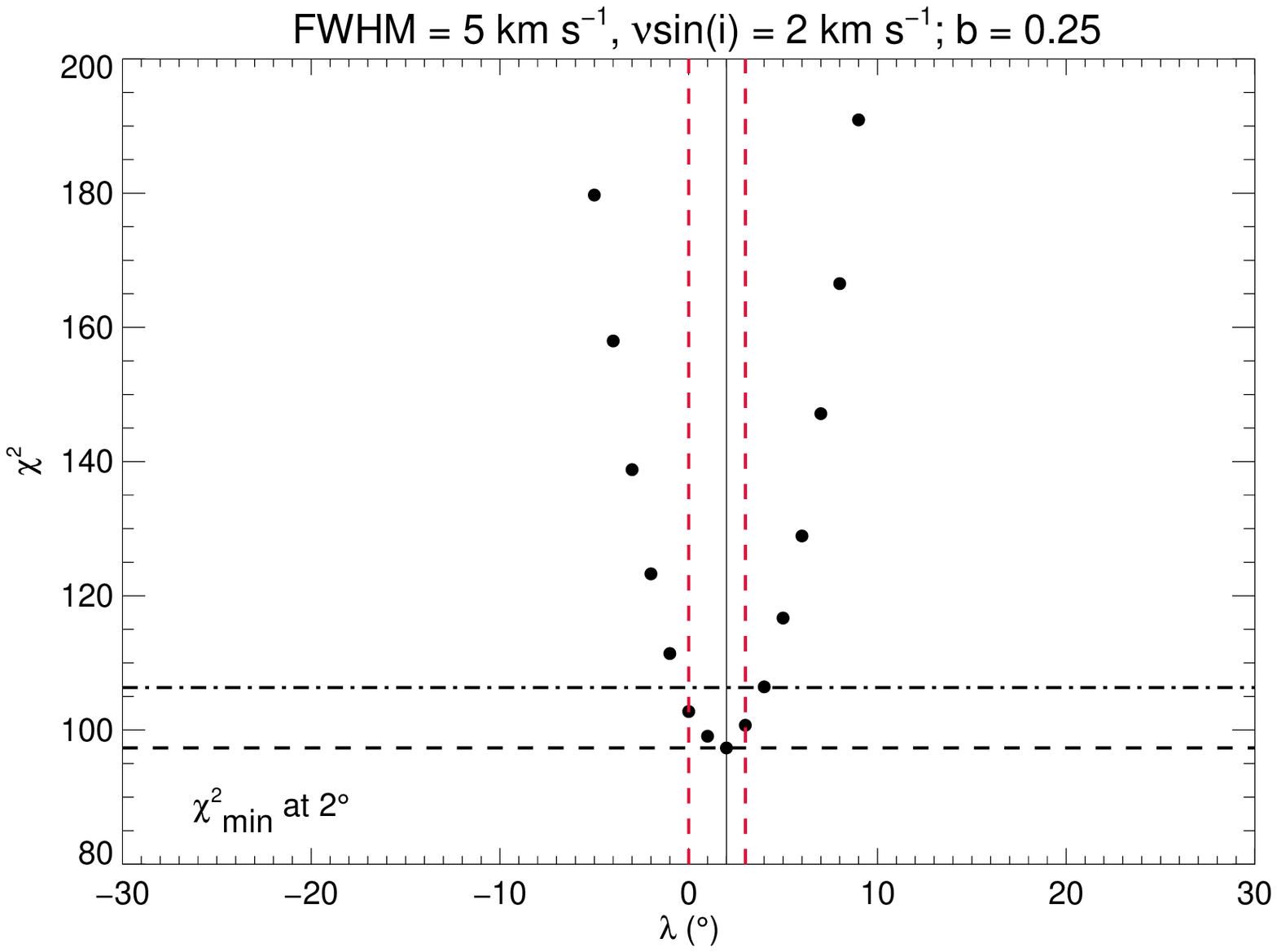}
\includegraphics[scale=0.44]{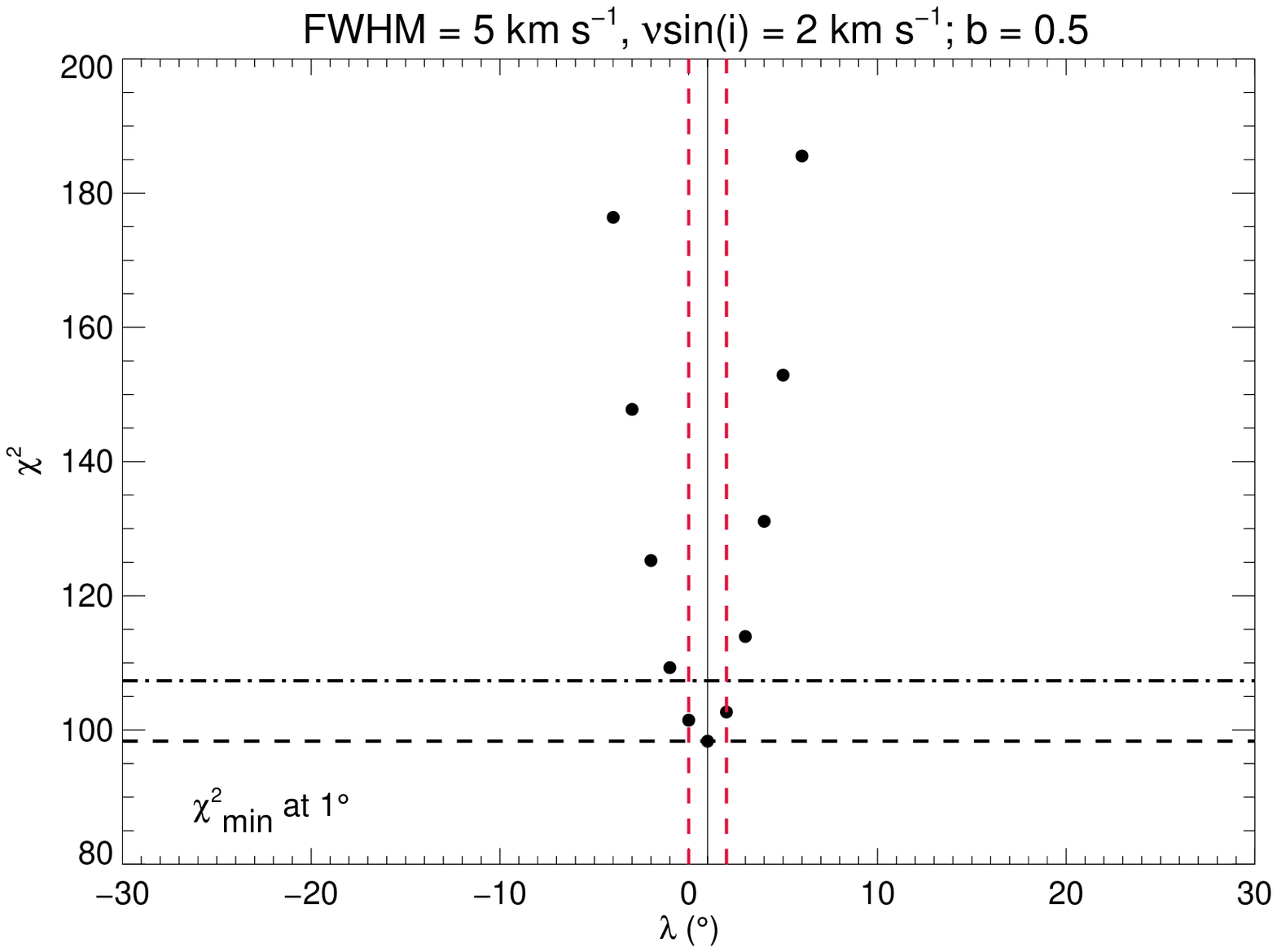}
\includegraphics[scale=0.44]{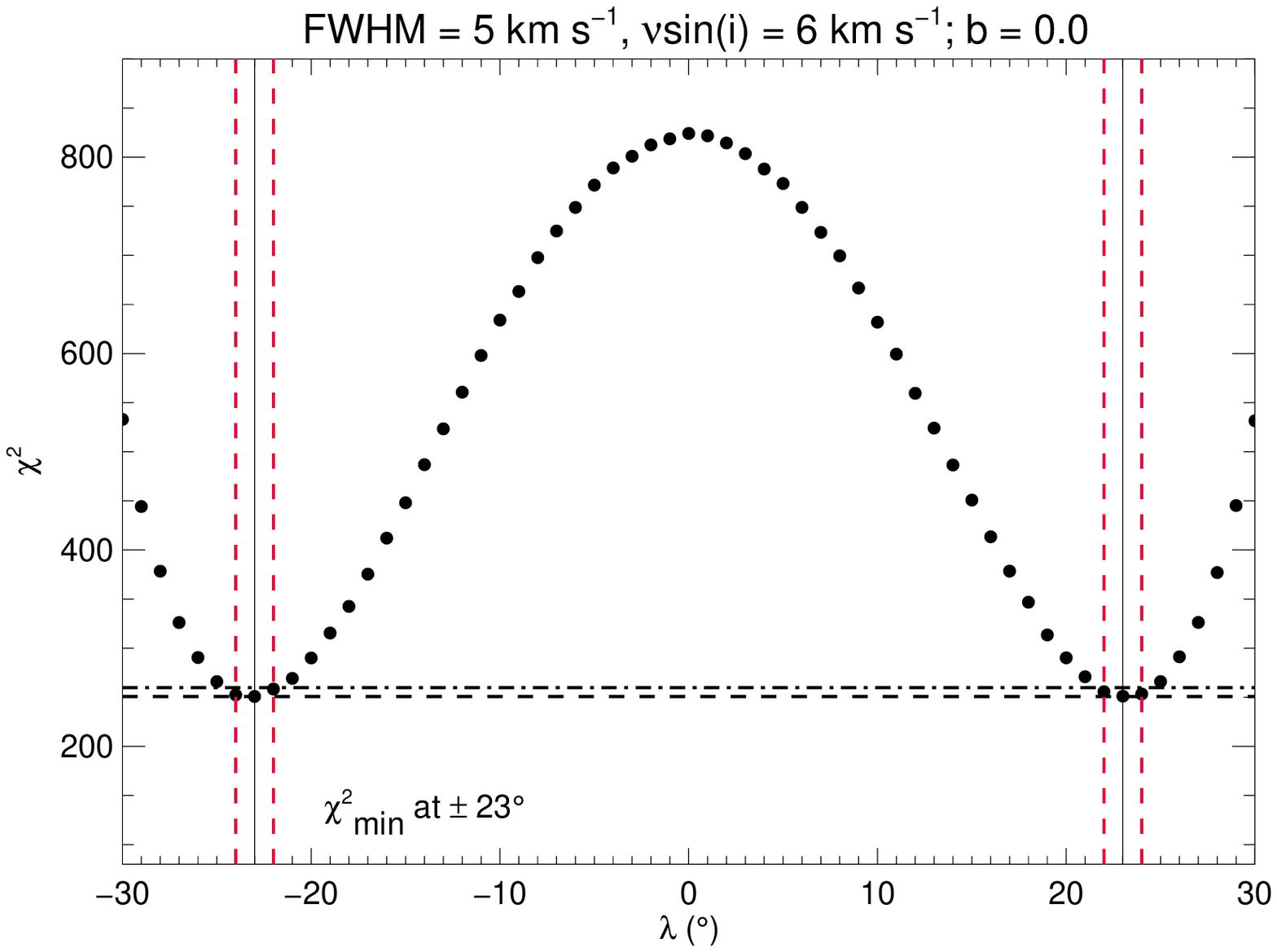}
\caption{$\chi^2$ maps for four different systems, using the C14 grid. The solid vertical lines indicate the $\chi^2$ minima, the horizontal dashed and dot-dashed lines represent the $\Delta \chi^2 = 9$ regions, and the vertical dashed red lines indicate the corresponding $\lambda$ limits that fall within $\Delta \chi^2 = 9$ (and therefore indicate a 3$\sigma$ confidence interval on the minimum $\chi^2$). Top and Bottom Left: illustrate the decrease in degeneracy between $\chi^2$ and $\lambda$ at increasing impact factor, in clockwise order (examples illustrated only for the Gaussian intrinsic profile scenario). Bottom Right: illustrates a double $\chi^2$ minimum found (example is for the single asymmetric intrinsic profile scenario).} 
\label{fig:chimaps}
\end{center}
\end{figure*}
\end{center}

If the true intrinsic profile is instead represented by a single (i.e. constant across the stellar disc) asymmetric profile, then for the slowly rotating star we can still recover $\lambda$ values that indicate spin-orbit alignment. Again the errors on $\lambda$ were much larger when b = 0 for the C14 comparison, but the fit was worse than when the true intrinsic profile was Gaussian. For the fast rotator with b=0, there were two local minima at $\lambda = \pm 23 \pm 1\degree$ for the C14 case and two local minima at $\lambda = -25$ and $+23^{+12}_{-5}$$\degree$ for the SOAP-T case (see bottom right of Figure~\ref{fig:chimaps} for illustration of the two local minima); the fit was also much worse with $\chi^2_r = 2.72$ and 2.79, respectively. Hence, for this case we could not recover the spin-orbit alignment when ignoring the CB effects. When b = 0.25 and 0.5, we were able to recover the spin-orbit alignment, but then the fits achieved a $\chi^2_r = 7.53$ and 6.56, respectively for the C14 case and 3.95 and 2.33, respectively, for the SOAP-T case. Note that given the degrees of freedom in this data set, according to a $\chi^2$ distribution there is a $<0.1$\% probability of achieving a $\chi^2_r > 1.8$, and therefore any fits with such a high $\chi^2_r$ should not be trusted.

\vspace{-10pt}

Finally, we considered the case when the true intrinsic profiles were represented by limb-dependent asymmetric profiles. In this case, the fits respond similarly to the previous case with the constant asymmetric profile: the errors on $\lambda$ were higher when b = 0 for the C14 case, and alignment was found for all impact factors for the slowly rotating star and also for the fast star when b~$\neq 0$. The main difference between considering a range of asymmetric profiles, as compared to a single (constant) asymmetric profile, was that the goodness of fit was significantly worse for the fast rotating star when comparing with the C14 grid (with $\chi^2_r = 4.41, 18.85,$ and 10.96 for b~=~0, 0.25, and 0.5, respectively). We note that such poor fits could cause observers to assume they have under-estimated their errors, even if they have in fact obtained the true obliquity. In turn, this may prompt a renormalisation of the errors to achieve a best fit $\chi^2_r$ closer to 1; in which case, some errors on $\lambda$ reported in the literature may actually be overestimated for faster rotators. 

In general, the C14 grid produced much larger error on $\lambda$ when b = 0, and also to a lesser extent when the star rotated slower. This is because the $\chi^2$-$\lambda$ distribution has a broad minima when b = 0 that narrows with higher impact factors (and is also slightly narrower for the faster star) -- see Figure~\ref{fig:chimaps} for examples. Hence, there is a degeneracy between the minimum $\chi^2$ and the recovered $\lambda$, at least for very low impact factors. This indicates a potential degeneracy between recoverable obliquities and the CB variation. However, the narrowing of the $\chi^2$-$\lambda$ distribution for higher impact factors indicates that this potential degeneracy may weaken when b $\neq$ 0. Note that we cannot conclude that a degeneracy between CB and $\lambda$ can be completely broken for non-zero impact factors as this would require us to explore a range of impact factors and star-planet systems, as well as allowing for additional effects such as differential rotation (all of which is beyond the scope of this paper, but will be pursued in forthcoming publications).

Overall, our results provide evidence that the presence of a variable CB may inflate errors on $\lambda$, at least for very low impact factors. Additionally, both stellar grids show that neglecting to model an asymmetric intrinsic profile is more important for fast rotators and may result in incorrect misalignment measurements and/or very poorly fit models (which may cause observers to overestimate their errors in an attempt to improve the fit).

\section{Summary and Concluding Remarks}
\label{sec:conc}
Throughout this paper, we go beyond the classical RM modelling by including the expected variations across the stellar disc in the both the net convective blueshift and the stellar photospheric profile shape. To study the impact of these variations we used two different stellar models, SOAP-T and the Sun-as-a-star grid from \cite{cegla14b}. We simulated the transit of an aligned hot Jupiter with a 4 d orbit, and explored a range of (solid body) stellar rotation rates and intrinsic profile widths and shapes. The convective centre-to-limb variation in the model stars was based off results from a 3D MHD solar surface simulation. The asymmetry/shape of the intrinsic profile, representing the stellar photosphere, was varied by injecting granulation line profiles synthesised from the aforementioned MHD simulation; note the simulated line profiles were taken from only one position in time as we wanted to isolate the centre-to-limb variations from any temporal variability (i.e. granular evolution). We also quantified the impact of these convective effects on the measured obliquity of this planetary system.

To quantify the impact on obliquity, we examined the best-fit (as determined by $\chi^2$ minimisation) between models without convection (but with a variety of obliquities) and models with convective centre-to-limb variations (and a variety of true intrinsic profile shapes, i.e. Gaussian, constant asymmetric, range of asymmetries). These tests were carried out for both a fast ($v \sin i$ = 6 km s$^{-1}$) and slowly ($v \sin i$ = 2 km s$^{-1}$) rotating star, and for systems with impact factors of 0, 0.25, and 0.5. 

The findings of our study are summarised below: 

\begin{itemize}

\item The presence of a centre-to-limb variation in the net convective blueshift produces an asymmetric disc-integrated profile, even if the local intrinsic line profiles are Gaussian. This is because limb darkening creates an uneven weighting across the \textit{radially} symmetric centre-to-limb velocity shifts (e.g. an annuli at disc centre has a different brightness and net RV shift than an annuli near the limb).

\item 
The RVs measured during transit should be the sum of an odd (stellar rotation) and even (convective variation) function. However, this is not reflected in the velocity centroid determined from the mean of a Gaussian fit because the profiles on the blueshifted hemisphere have a different asymmetry to those on the redshifted hemisphere (due to the interplay of the rotation and convection). Hence, the residuals between models with and without convection are slightly asymmetric. 

\item The shape and amplitude of the residuals between RM curves with and without a centre-to-limb convective variation depend on the star's $v \sin i$ and intrinsic profile FWHM. The amplitude of the residuals increase with increasing $v \sin i$, and decreasing FWHM. We believe this unexpected behaviour could be related to two phenomena. First, fitting a Gaussian to an asymmetric profile produces offsets from the true velocity centroid, and these offsets/errors increase with increasing $v \sin i$ and decreasing FWHM. Second, it may be caused by the increased contribution from the limb to the disc-integrated profile at greater $v \sin i$ \citep[][and references therein]{bruning90,smith87}, where the net CB is most redshifted.

\item 
When the $v \sin i$ of the star is less then the FWHM of the intrinsic profile, the residuals between a model star with and without a centre-to-limb convective variation results in a blueshift at ingress and egress (where the obscured convective velocities are most redshifted) and a redshift near mid-transit (where the convective velocities are most blueshifted). However, if the $v \sin i$ of the star is greater than the FWHM of the intrinsic profile, then the ingress and egress are also redshifted; the reason for this behaviour is not clear, but it may also be related to the RV fitting procedure and/or the increased contribution from the net CB at the limb once the $v \sin i$ is greater than the intrinsic broadening \citep{gray85}.

\item 
The amplitude of the residuals between stars with and without centre-to-limb convective variations also depends on the correct modelling of the intrinsic line profile shapes. For slow rotators, $v \sin i \le$~2~km~s$^{-1}$, the impact of the convective blueshift contribution can be seen in the residuals, and dominates over the intrinsic profile modelling, with amplitudes $<$ 0.5~m~s$^{-1}$. While these effects may be negligible now, this is unlikely to be the case once spectrographs reach 10~cm~s$^{-1}$ precision. For faster rotators, 3~km~s$^{-1}$~$ \le v \sin i \le $~10~km~s$^{-1}$, an incorrect modelling of the intrinsic profile shape dominates the residuals. If the true intrinsic profile can be represented by one constant, asymmetric profile (but is incorrectly modelled by a Gaussian), the residuals ranged from $\sim$ 1 to 4 m~s$^{-1}$, but if the asymmetries changed across the stellar disc then the residuals ranged from $\sim$ 0.5 to 9 m~s$^{-1}$ (with greater residuals for greater $v \sin i$). The exact amplitude of the residuals will depend on the convective properties of the star and the level of asymmetry of the observed intrinsic line profile/CCF, and therefore may be greater or less than found here. 

\item 
For a hot Jupiter with a 4 d orbit about a Sun-like star, neglecting to account for the centre-to-limb variation in convective blueshift led to an uncertainty in the obliquity of $\sim$10-20$\degree$ for aligned systems with an impact factor of 0. We believe this is due to a potential degeneracy between the projected obliquity, $\lambda$, and the convective blueshift. The uncertainty on the obliquity may decrease for non-zero impact factors, down to 1-3$\degree$. However, we cannot claim that such a degeneracy is completely broken as this was not found with both stellar grids; and also because we have only tested one aligned system under the assumption of solid body rotation (ignoring granular evolution and other contributions to the observed RVs). We also found that neglecting to properly model an asymmetric intrinsic profile may result in incorrect misalignment measurements for fast rotators (off by $\sim$ 20$\degree$ from the true projected obliquity). Additionally, incorrectly modelling the intrinsic profile shape also produced worse model fits, especially for the faster rotating stars which had `best-fit' models with extremely unlikely probabilities ($< 0.1$\%).

\end{itemize}

In this paper, we have found that the convective centre-to-limb variations in the stellar photosphere of a Sun-like star have the potential to significantly affect the RM waveform, even for the transit of a hot Jupiter. Not only can these variations lead to residuals on the m~s$^{-1}$ level, but if unaccounted for can also lead to both incorrect (projected) obliquity measurements and incorrect error estimations on the (projected) obliquity.  

The residuals predicted between observed data and traditional RM models (that ignore the centre-to-limb variation in convection) should be measurable by current spectrographs if the $v \sin i$ is greater than $\sim$ 3~km~s$^{-1}$ and if the shape of the intrinsic profile/CCF is non-Gaussian. Herein, we have shown that these residuals increase with increasing $v \sin i$ and decreasing intrinsic profile FWHM. Furthermore, these effects may even be able to explain some of the correlated residuals reported in the literature between observed transits and previous RM models (e.g. those found for HD 189733;\citealt{winn06, triaud09}  -- note, this is in agreement with the hypothesis put forward by \citealt{czesla15} in their study of the centre-to-limb intensity variations). 

In forthcoming publications, we aim to search for these effects observationally and also to predict them for a variety of star-planet systems (e.g. with varying obliquity, planet mass/radius/separation, impact factors, stellar rotation, and spectral type/magnetic field strength). Of particular importance is to quantify the convective contribution to the observed RM signal for small planets, as it may completely dominate over the contribution from stellar rotation (especially for slow rotators), and to account for temporal variations from granular evolution. 

As instrumental precision increases it is ever more important to correctly account for the contribution from the stellar surface in the observed RVs of high precision transit measurements. Our results indicate that neglecting to do say may hamper and/or bias our interpretation of planetary evolution and migration. Fortunately, some of the residuals from failing to account for convection in the observed RM waveform should be readily detectable, and therefore may help confirm the proper way to include the convective effects in future RM modelling.

\section*{\sc Acknowledgments}
We thank the anonymous referee for their considered report, that led to a much clearer and more concise manuscript, and provided important insight into the behaviour of the residuals. The authors would also like to thank E. de Mooij for useful discussions that improved computational speed. HMC and CAW gratefully acknowledge support from the Leverhulme Trust (grant RPG-249). CAW also acknowledges support from STFC grant ST/L000709/1. MO acknowledges support by the Centro de
Astrof\'{i}sca da Universidade do Porto through grant CAUP-15/2014-BDP. This work was supported by Funda\c{c}\~ao para a Ci\^encia e a Tecnologia (FCT) through the research grants UID/FIS/04434/2013 and PTDC/FIS-AST/1526/2014. PF and NCS also acknowledge the support from FCT through Investigador FCT contracts of reference IF/01037/2013 and IF/00169/2012, respectively, and POPH/FSE (EC) by FEDER funding through the program ``Programa Operacional de Factores de Competitividade - COMPETE''. PF further acknowledges support from Funda\c{c}\~ao para a Ci\^encia e a Tecnologia (FCT) in the form of an exploratory project of reference IF/01037/2013CP1191/CT0001. SS is the recipient of an Australian Research Council’s Future Fellowship (project number FT120100057). This research has made use of NASA's Astrophysics Data System Bibliographic Services.

\bibliographystyle{apj}
\bibliography{abbrev,mybib}

\begin{thebibliography}{40}
\expandafter\ifx\csname natexlab\endcsname\relax\def\natexlab#1{#1}\fi

\bibitem[{{Beckers}(2007)}]{beckers07}
{Beckers}, J.~M. 2007, Astronomische Nachrichten, 328, 1084

\bibitem[{{Boisse} {et~al.}(2011){Boisse}, {Bouchy}, {H{\'e}brard}, {Bonfils},
  {Santos}, \& {Vauclair}}]{boisse11}
{Boisse}, I., {Bouchy}, F., {H{\'e}brard}, G., {et~al.} 2011, A\&A, 528, A4+

\bibitem[{{Bou{\'e}} {et~al.}(2013){Bou{\'e}}, {Montalto}, {Boisse}, {Oshagh},
  \& {Santos}}]{boue13}
{Bou{\'e}}, G., {Montalto}, M., {Boisse}, I., {Oshagh}, M., \& {Santos}, N.~C.
  2013, \aap, 550, A53

\bibitem[{{Bruning} \& {Saar}(1990)}]{bruning90}
{Bruning}, D.~H., \& {Saar}, S.~H. 1990, in Astronomical Society of the Pacific
  Conference Series, Vol.~9, Cool Stars, Stellar Systems, and the Sun, ed.
  G.~{Wallerstein}, 165--167

\bibitem[{{Cegla} {et~al.}(2013){Cegla}, {Shelyag}, {Watson}, \&
  {Mathioudakis}}]{cegla13}
{Cegla}, H.~M., {Shelyag}, S., {Watson}, C.~A., \& {Mathioudakis}, M. 2013,
  \apj, 763, 95

\bibitem[{{Cegla} {et~al.}(2014){Cegla}, {Watson}, {Shelyag}, \&
  {Mathioudakis}}]{cegla14b}
{Cegla}, H.~M., {Watson}, C.~A., {Shelyag}, S., \& {Mathioudakis}, M. 2014,
  ArXiv:1408.2301

\bibitem[{{Cegla} {et~al.}(2015){Cegla}, {Watson}, {Shelyag}, {Mathioudakis},
  \& {Moutari}}]{cegla15c}
{Cegla}, H.~M., {Watson}, C.~A., {Shelyag}, S., {Mathioudakis}, M., \&
  {Moutari}, S. 2015, in prep

\bibitem[{{Cegla} {et~al.}(2012){Cegla}, {Watson}, {Marsh}, {Shelyag},
  {Moulds}, {Littlefair}, {Mathioudakis}, {Pollacco}, \& {Bonfils}}]{cegla12}
{Cegla}, H.~M., {Watson}, C.~A., {Marsh}, T.~R., {et~al.} 2012, \mnras, 421,
  L54

\bibitem[{{Collier Cameron} {et~al.}(2010){Collier Cameron}, {Bruce}, {Miller},
  {Triaud}, \& {Queloz}}]{cameron10}
{Collier Cameron}, A., {Bruce}, V.~A., {Miller}, G.~R.~M., {Triaud},
  A.~H.~M.~J., \& {Queloz}, D. 2010, \mnras, 403, 151

\bibitem[{{Czesla} {et~al.}(2015){Czesla}, {Klocov{\'a}}, {Khalafinejad},
  {Wolter}, \& {Schmitt}}]{czesla15}
{Czesla}, S., {Klocov{\'a}}, T., {Khalafinejad}, S., {Wolter}, U., \&
  {Schmitt}, J.~H.~M.~M. 2015, \aap, 582, A51

\bibitem[{{Desidera} {et~al.}(2004){Desidera}, {Gratton}, {Endl}, {Claudi},
  {Cosentino}, {Barbieri}, {Bonanno}, {Lucatello}, {Martinez Fiorenzano},
  {Marzari}, \& {Scuderi}}]{desidera04}
{Desidera}, S., {Gratton}, R.~G., {Endl}, M., {et~al.} 2004, A\&A, 420, L27

\bibitem[{{Dravins}(1987)}]{dravins87a}
{Dravins}, D. 1987, \aap, 172, 200

\bibitem[{{Dravins} {et~al.}(2015){Dravins}, {Ludwig}, {Dahlen}, \&
  {Pazira}}]{dravins15}
{Dravins}, D., {Ludwig}, H.-G., {Dahlen}, E., \& {Pazira}, H. 2015, in
  Cambridge Workshop on Cool Stars, Stellar Systems, and the Sun, Vol.~18, 18th
  Cambridge Workshop on Cool Stars, Stellar Systems, and the Sun, ed. G.~T.
  {van Belle} \& H.~C. {Harris}, 853--868

\bibitem[{{Dravins} \& {Nordlund}(1990)}]{dravins90b}
{Dravins}, D., \& {Nordlund}, A. 1990, \aap, 228, 203

\bibitem[{{Dumusque} {et~al.}(2011{\natexlab{a}}){Dumusque}, {Santos}, {Udry},
  {Lovis}, \& {Bonfils}}]{dumusque11b}
{Dumusque}, X., {Santos}, N.~C., {Udry}, S., {Lovis}, C., \& {Bonfils}, X.
  2011{\natexlab{a}}, A\&A, 527, A82+

\bibitem[{{Dumusque} {et~al.}(2011{\natexlab{b}}){Dumusque}, {Udry}, {Lovis},
  {Santos}, \& {Monteiro}}]{dumusque11a}
{Dumusque}, X., {Udry}, S., {Lovis}, C., {Santos}, N.~C., \& {Monteiro},
  M.~J.~P.~F.~G. 2011{\natexlab{b}}, A\&A, 525, A140+

\bibitem[{{Figueira} {et~al.}(2010){Figueira}, {Marmier}, {Bonfils}, {di
  Folco}, {Udry}, {Santos}, {Lovis}, {M{\'e}gevand}, {Melo}, {Pepe}, {Queloz},
  {S{\'e}gransan}, {Triaud}, \& {Viana Almeida}}]{figueria10}
{Figueira}, P., {Marmier}, M., {Bonfils}, X., {et~al.} 2010, A\&A, 513, L8+

\bibitem[{{Gray}(2005)}]{gray05}
{Gray}, D.~F. 2005, {The Observation and Analysis of Stellar Photospheres}

\bibitem[{{Gray} \& {Toner}(1985)}]{gray85}
{Gray}, D.~F., \& {Toner}, C.~G. 1985, \pasp, 97, 543

\bibitem[{{Hirano} {et~al.}(2010){Hirano}, {Suto}, {Taruya}, {Narita}, {Sato},
  {Johnson}, \& {Winn}}]{hirano10}
{Hirano}, T., {Suto}, Y., {Taruya}, A., {et~al.} 2010, \apj, 709, 458

\bibitem[{{Hu{\'e}lamo} {et~al.}(2008){Hu{\'e}lamo}, {Figueira}, {Bonfils},
  {Santos}, {Pepe}, {Gillon}, {Azevedo}, {Barman}, {Fern{\'a}ndez}, {di Folco},
  {Guenther}, {Lovis}, {Melo}, {Queloz}, \& {Udry}}]{huelamo08}
{Hu{\'e}lamo}, N., {Figueira}, P., {Bonfils}, X., {et~al.} 2008, A\&A, 489, L9

\bibitem[{{McLaughlin}(1924)}]{mclaughlin}
{McLaughlin}, D.~B. 1924, ApJ, 60, 22

\bibitem[{{Meunier} \& {Lagrange}(2013)}]{meunier13}
{Meunier}, N., \& {Lagrange}, A.-M. 2013, \aap, 551, A101

\bibitem[{{Miller} {et~al.}(2010){Miller}, {Collier Cameron}, {Simpson},
  {Pollacco}, {Enoch}, {Gibson}, {Queloz}, {Triaud}, {H{\'e}brard}, {Boisse},
  {Moutou}, \& {Skillen}}]{miller10}
{Miller}, G.~R.~M., {Collier Cameron}, A., {Simpson}, E.~K., {et~al.} 2010,
  \aap, 523, A52

\bibitem[{{Oshagh} {et~al.}(2013{\natexlab{a}}){Oshagh}, {Boisse}, {Bou{\'e}},
  {Montalto}, {Santos}, {Bonfils}, \& {Haghighipour}}]{oshagh13}
{Oshagh}, M., {Boisse}, I., {Bou{\'e}}, G., {et~al.} 2013{\natexlab{a}}, \aap,
  549, A35

\bibitem[{{Oshagh} {et~al.}(2013{\natexlab{b}}){Oshagh}, {Bou{\'e}},
  {Figueira}, {Santos}, \& {Haghighipour}}]{oshagh14}
{Oshagh}, M., {Bou{\'e}}, G., {Figueira}, P., {Santos}, N.~C., \&
  {Haghighipour}, N. 2013{\natexlab{b}}, \aap, 558, A65

\bibitem[{{Pepe} {et~al.}(2014){Pepe}, {Molaro}, {Cristiani}, {Rebolo},
  {Santos}, {Dekker}, {M{\'e}gevand}, {Zerbi}, {Cabral}, {Di Marcantonio},
  {Abreu}, {Affolter}, {Aliverti}, {Allende Prieto}, {Amate}, {Avila},
  {Baldini}, {Bristow}, {Broeg}, {Cirami}, {Coelho}, {Conconi}, {Coretti},
  {Cupani}, {D'Odorico}, {De Caprio}, {Delabre}, {Dorn}, {Figueira}, {Fragoso},
  {Galeotta}, {Genolet}, {Gomes}, {Gonz{\'a}lez Hern{\'a}ndez}, {Hughes},
  {Iwert}, {Kerber}, {Landoni}, {Lizon}, {Lovis}, {Maire}, {Mannetta},
  {Martins}, {Monteiro}, {Oliveira}, {Poretti}, {Rasilla}, {Riva}, {Santana
  Tschudi}, {Santos}, {Sosnowska}, {Sousa}, {Span{\'o}}, {Tenegi}, {Toso},
  {Vanzella}, {Viel}, \& {Zapatero Osorio}}]{pepe14}
{Pepe}, F., {Molaro}, P., {Cristiani}, S., {et~al.} 2014, ArXiv e-prints

\bibitem[{{Queloz} {et~al.}(2001){Queloz}, {Henry}, {Sivan}, {Baliunas},
  {Beuzit}, {Donahue}, {Mayor}, {Naef}, {Perrier}, \& {Udry}}]{queloz01}
{Queloz}, D., {Henry}, G.~W., {Sivan}, J.~P., {et~al.} 2001, A\&A, 379, 279

\bibitem[{{Robertson} {et~al.}(2015){Robertson}, {Roy}, \&
  {Mahadevan}}]{robertson15}
{Robertson}, P., {Roy}, A., \& {Mahadevan}, S. 2015, \apjl, 805, L22

\bibitem[{{Rossiter}(1924)}]{rossiter}
{Rossiter}, R.~A. 1924, ApJ, 60, 15

\bibitem[{{Saar} \& {Donahue}(1997)}]{saar97}
{Saar}, S.~H., \& {Donahue}, R.~A. 1997, ApJ, 485, 319

\bibitem[{{Santos} {et~al.}(2014){Santos}, {Mortier}, {Faria}, {Dumusque},
  {Adibekyan}, {Delgado-Mena}, {Figueira}, {Benamati}, {Boisse}, {Cunha},
  {Gomes da Silva}, {Lo Curto}, {Lovis}, {Martins}, {Mayor}, {Melo}, {Oshagh},
  {Pepe}, {Queloz}, {Santerne}, {S{\'e}gransan}, {Sozzetti}, {Sousa}, \&
  {Udry}}]{santos14}
{Santos}, N.~C., {Mortier}, A., {Faria}, J.~P., {et~al.} 2014, \aap, 566, A35

\bibitem[{{Schrijver} \& {Zwaan}(2000)}]{schrijver00}
{Schrijver}, C.~J., \& {Zwaan}, C. 2000, {Solar and Stellar Magnetic Activity}
  (Cambridge University Press)

\bibitem[{{Shporer} \& {Brown}(2011)}]{shporer11}
{Shporer}, A., \& {Brown}, T. 2011, \apj, 733, 30

\bibitem[{{Smith} {et~al.}(1987){Smith}, {Livingston}, \& {Huang}}]{smith87}
{Smith}, M.~A., {Livingston}, W., \& {Huang}, Y.-R. 1987, \pasp, 99, 297

\bibitem[{{Triaud} {et~al.}(2009){Triaud}, {Queloz}, {Bouchy}, {Moutou},
  {Collier Cameron}, {Claret}, {Barge}, {Benz}, {Deleuil}, {Guillot},
  {H{\'e}brard}, {Lecavelier Des {\'E}tangs}, {Lovis}, {Mayor}, {Pepe}, \&
  {Udry}}]{triaud09}
{Triaud}, A.~H.~M.~J., {Queloz}, D., {Bouchy}, F., {et~al.} 2009, \aap, 506,
  377

\bibitem[{{Trujillo Bueno} {et~al.}(2004){Trujillo Bueno}, {Shchukina}, \&
  {Asensio Ramos}}]{bueno04}
{Trujillo Bueno}, J., {Shchukina}, N., \& {Asensio Ramos}, A. 2004, \nat, 430,
  326

\bibitem[{{V{\"o}gler} {et~al.}(2005){V{\"o}gler}, {Shelyag}, {Sch{\"u}ssler},
  {Cattaneo}, {Emonet}, \& {Linde}}]{vogler05}
{V{\"o}gler}, A., {Shelyag}, S., {Sch{\"u}ssler}, M., {et~al.} 2005, A\&A, 429,
  335

\bibitem[{{Winn}(2007)}]{winn07}
{Winn}, J.~N. 2007, in Astronomical Society of the Pacific Conference Series,
  Vol. 366, Transiting Extrapolar Planets Workshop, ed. C.~{Afonso},
  D.~{Weldrake}, \& T.~{Henning}, 170

\bibitem[{{Winn} {et~al.}(2006){Winn}, {Johnson}, {Marcy}, {Butler}, {Vogt},
  {Henry}, {Roussanova}, {Holman}, {Enya}, {Narita}, {Suto}, \&
  {Turner}}]{winn06}
{Winn}, J.~N., {Johnson}, J.~A., {Marcy}, G.~W., {et~al.} 2006, \apjl, 653, L69

\end{thebibliography}

\begin{appendix}
\section{Comparing C14 grid and SOAP-T}
\label{appen:compare}
\begin{center}
\begin{figure*}[t!]
\begin{center}
\includegraphics[scale=0.44]{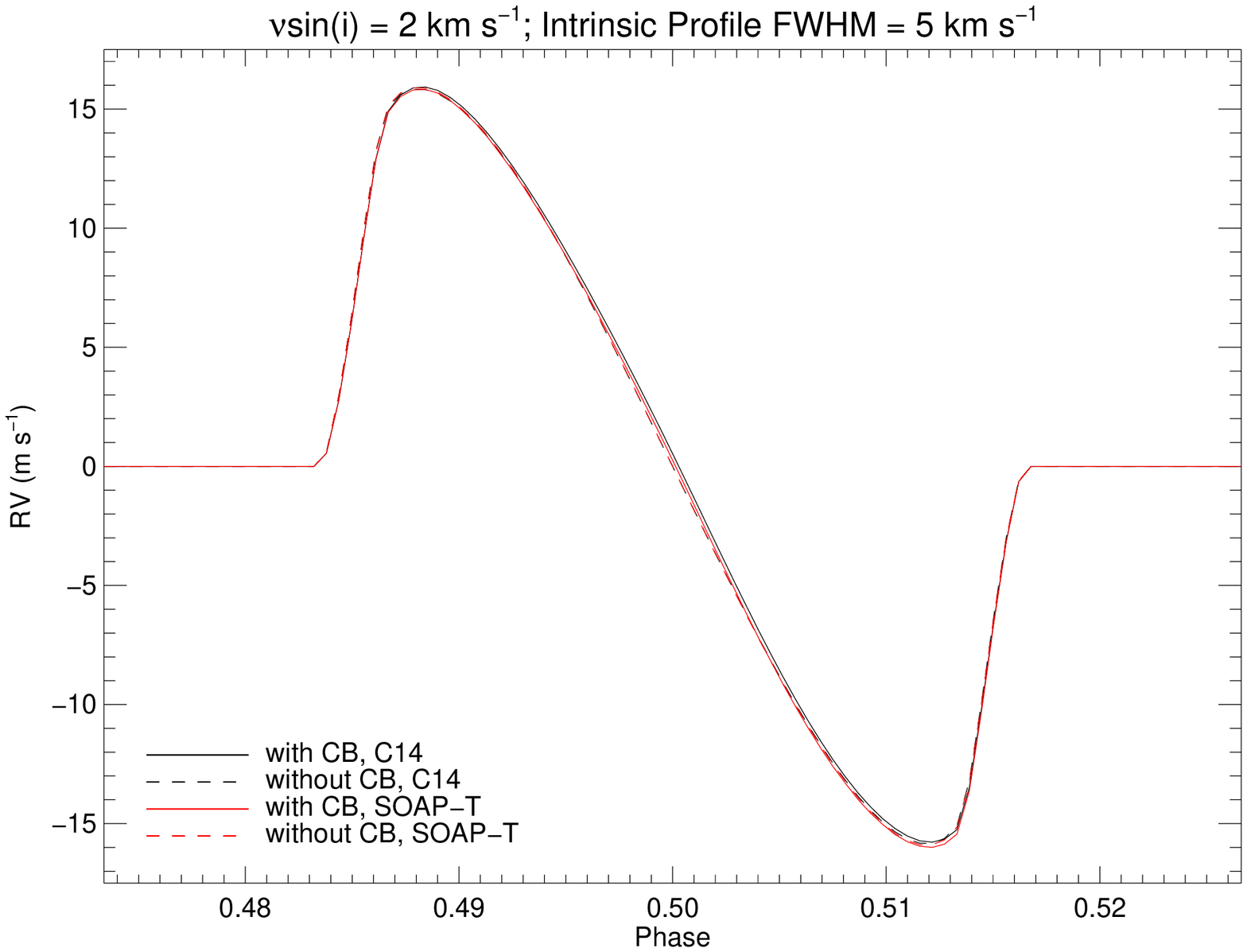}
\includegraphics[scale=0.44]{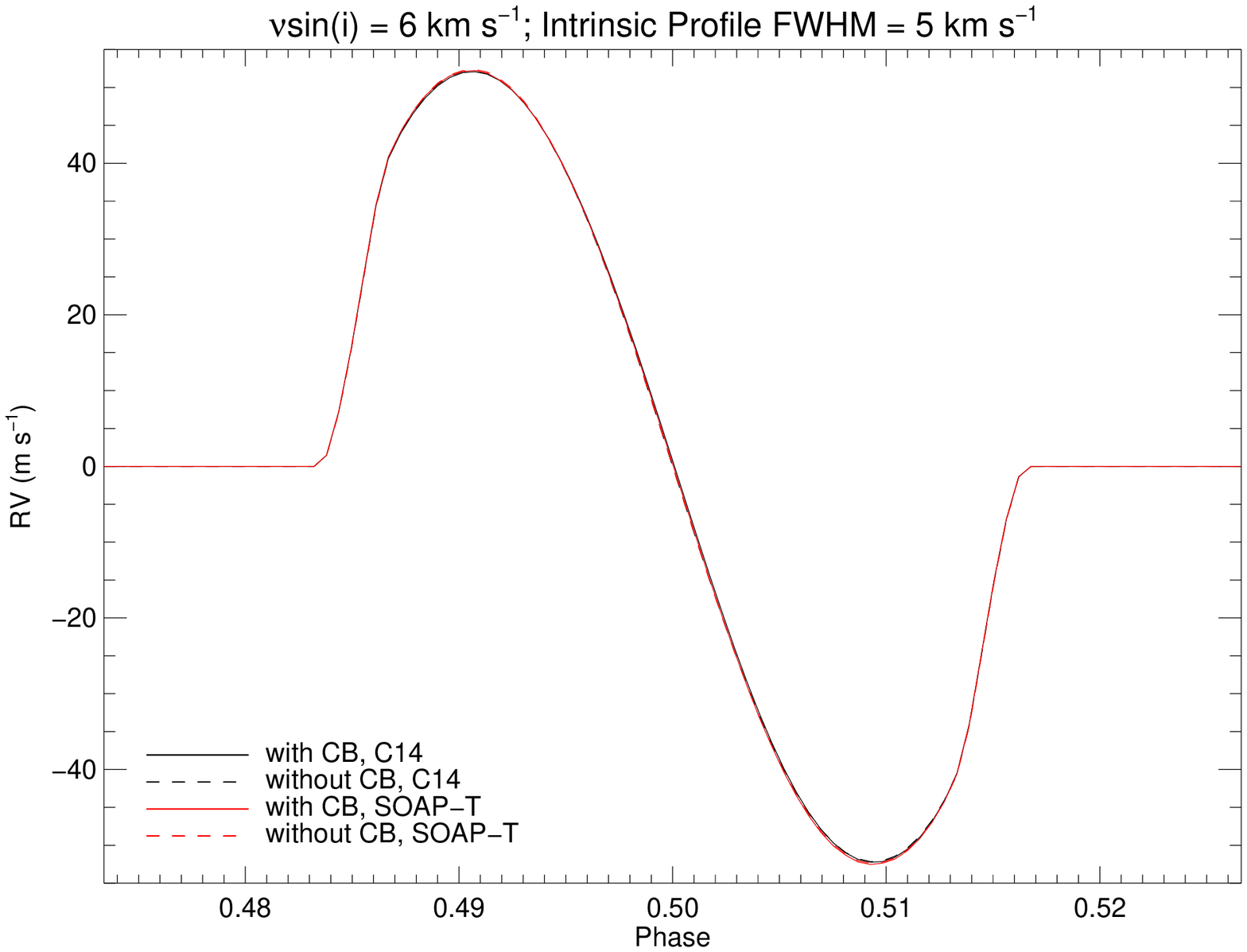}
\includegraphics[scale=0.44]{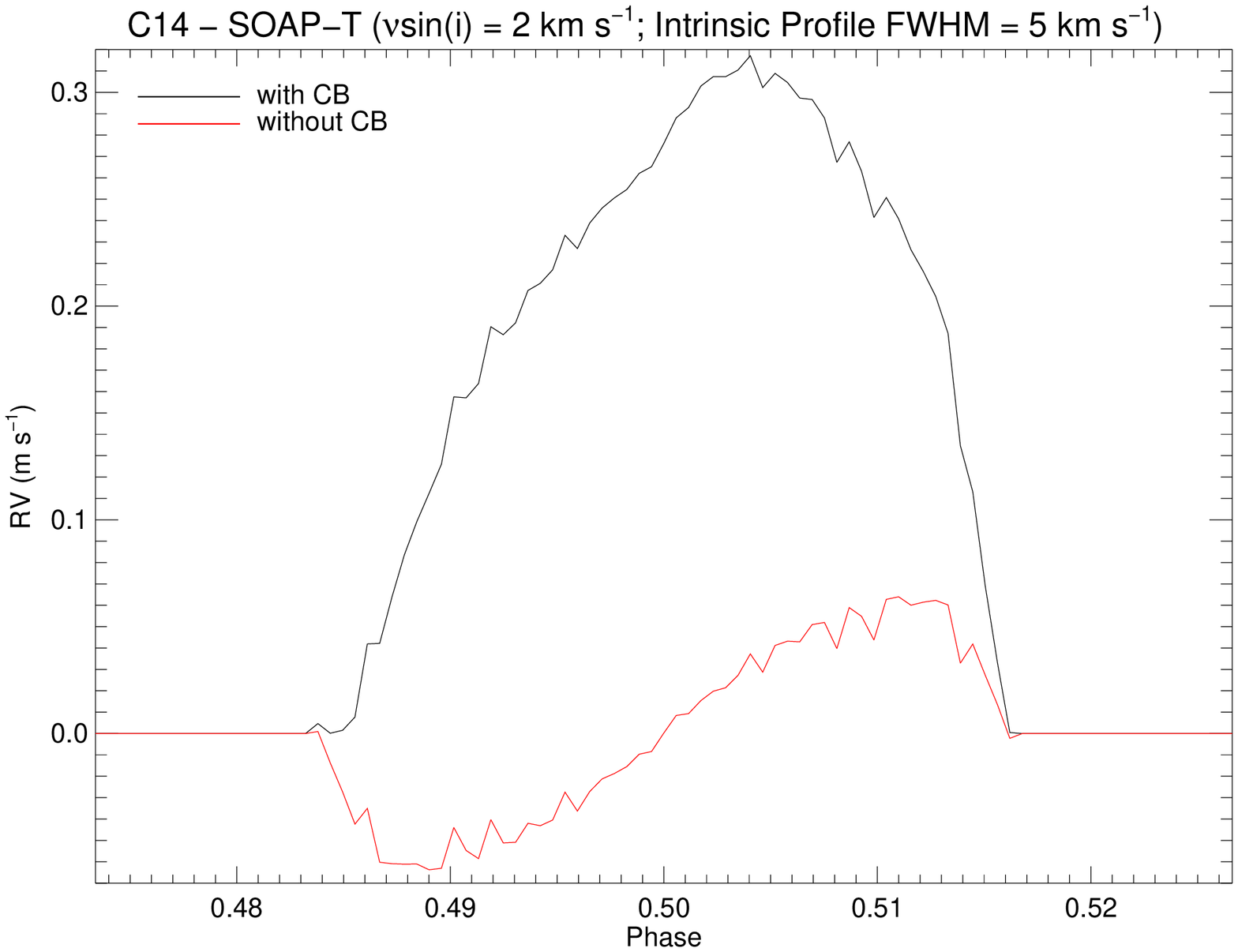}
\includegraphics[scale=0.44]{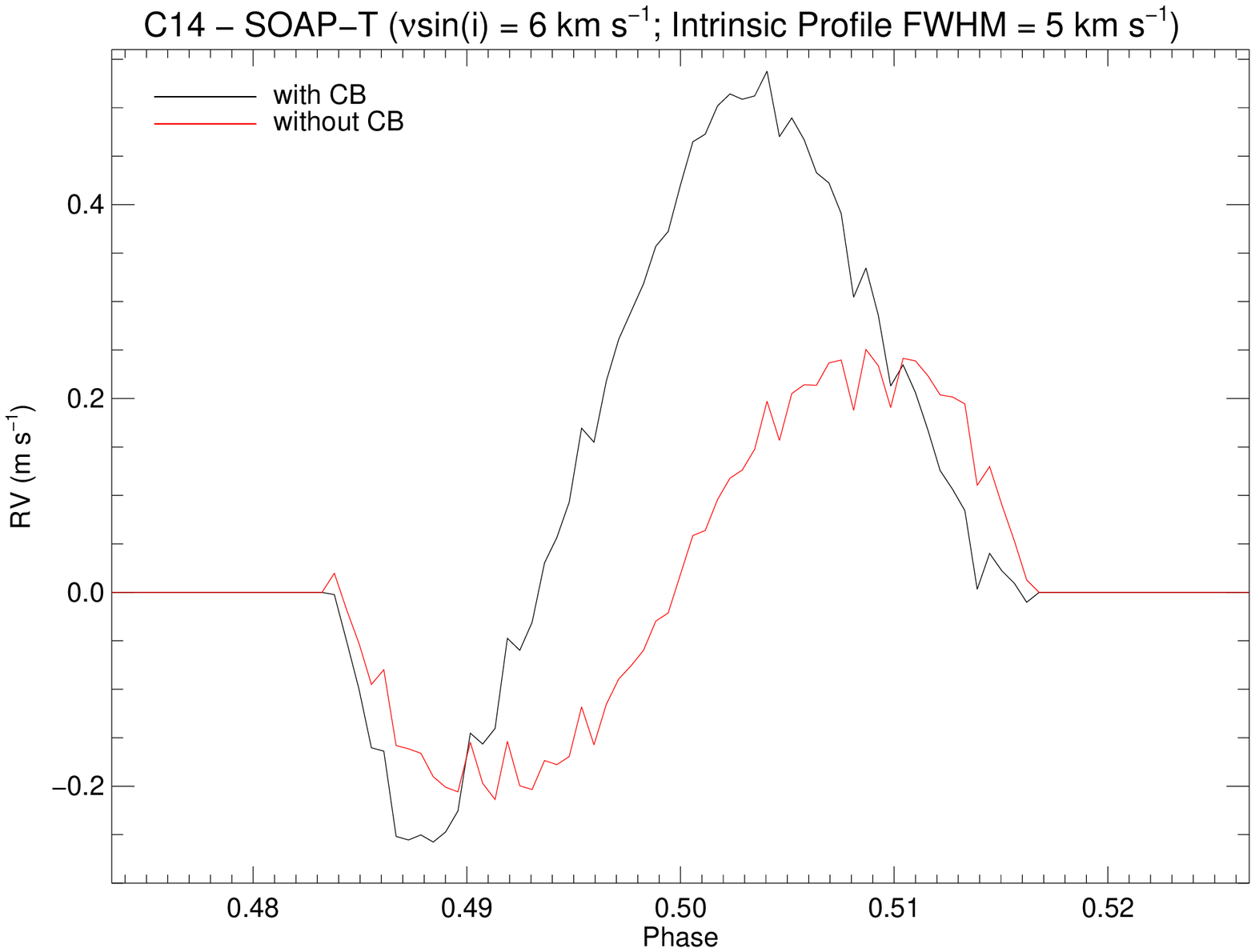}
\caption{Top: RM curves for both the C14 grid (black) and SOAP-T (red), for observations with (dashed) and without (solid) CB, for two different systems: $v \sin i$ = 2 km s~$^{-1}$ (left) and $v \sin i$ = 6 km s~$^{-1}$ (right); the injected intrinsic profiles in each case had a FWHM = 5 km~s$^{-1}$. Bottom: The residuals between the C14 and SOAP-T RM curves for observations with (black) and without (red) CB, for both systems. Note that the inclusion of the CB led to non-zero out-of-transit RVs for both models, which were subtracted from the RM curves as we are interested in the relative RV variations. 
} 
\label{fig:compsoap}
\end{center}
\end{figure*}
\end{center}

Since we used two independent stellar models throughout the paper it is important to examine the differences between the resultant RM curves. To do so, we inspected the residuals between the RM curves produced by each model (i.e. C14 - SOAP-T), both with and without the CB variation. These residuals, for two systems, are shown in Figure~\ref{fig:compsoap}. One system has $v \sin i$ = 2 km~s$^{-1}$ and the other has $v \sin i$ = 6 km~s$^{-1}$, both have Gaussian profiles injected with a FWHM = 5 km~s$^{-1}$. We examined these systems in case the residuals depended on the relationship between the $v \sin i$ and the injected profile FWHM (since the average Gaussian fit to a CCF depends on both these quantities, it is possible the aforementioned residuals may also be impacted; \citealt{hirano10, boue13}). 

\vspace{-15pt}

The residuals between the two models for observations without CB (red curves in Figure~\ref{fig:compsoap}) show there is a small mismatch ($< \sim$10 cm~s$^{-1}$) throughout the transits between the two models. This mismatch is slightly larger for the observations with a larger $v \sin i$ (up to $\sim$20 cm~s$^{-1}$). Given their behaviour, we believe this is due to the differences in tiling between the two grids, which ultimately leads to a slightly different set of stellar rotational shifts. The residuals for observations with CB (black curves in Figure~\ref{fig:compsoap}) also show a small mismatch ($< \sim$50 cm~s$^{-1}$) throughout the transit, but the shape of this curve for both systems is peculiar. The exact cause of these strangely shaped residual curves is not clear, but we believe this is a reflection of the relationship between stellar grid tiling, injected profile shifts, and injected profile flux. Note that during this comparison we tried a variety of tiling changes designed to increase precision within each model, but none of these significantly altered the residual shapes/amplitudes.

We note that the residuals between the two codes are on the same order of magnitude as the RM waveform due to a variable CB across the disc. Consequently, these software packages could produce different results throughout the paper. However, this is unlikely as the residuals between the two codes have very different shapes compared to the RM waveform from CB alone. Moreover, the two codes are treated independently throughout the paper; the only exception to this is in Section~\ref{sec:spinorbit} when SOAP-T was used to fit the RM signals produced from the C14 grid when an asymmetric intrinsic profile was considered (this was because at present SOAP-T cannot handle non-Gaussian profiles). Hence, we do not expect the differences between the two software packages to impact our analysis of the results.
\end{appendix}

\end{document}